\documentclass{JINST}
\usepackage[english]{babel}
\usepackage{amsfonts}
\usepackage{makeidx}
\usepackage{amsmath}
\usepackage{cite}
\usepackage{endnotes}

\newcommand{\gsim}{~{\buildrel > \over {_\sim}}~}
\newcommand{\sqrtsNN}{\sqrt{s_{\scriptscriptstyle{{\rm NN}}}}}

\newcommand{\gev}{\mathrm{GeV}}
\newcommand{\tev}{\mathrm{TeV}}
\newcommand{\fm}{\mathrm{fm}}
\newcommand{\mm}{\mathrm{mm}}
\newcommand{\cm}{\mathrm{cm}}

\newcommand{\mum}{\mathrm{\mu m}}

\newcommand{\pt}{p_{\rm t}}

\newcommand{\dxloc}{\Delta x_{\rm loc}} 
\newcommand{\dxy}{\Delta xy\vert_{y=0}} 
\newcommand{\inst}[1]{$^{#1}$}
\title{Alignment of the ALICE Inner Tracking System with cosmic-ray tracks}

\author{
ALICE collaboration
} 

\abstract{ALICE (A Large Ion Collider Experiment) is the LHC (Large Hadron
Collider) experiment devoted to investigating the strongly interacting 
matter created in nucleus-nucleus collisions at the LHC energies.
The ALICE ITS, Inner Tracking System, consists of six cylindrical
layers of silicon detectors with three different technologies;
in the outward direction: 
two layers of pixel detectors, two layers each of drift, and strip detectors. 
The number of parameters to be determined in the spatial alignment of the 2198
sensor modules of the ITS is
about 13,000. The target alignment precision is 
 well below 10~$\mum$ in some cases (pixels). 
The sources of alignment information include survey measurements,
and the reconstructed tracks from cosmic rays and from proton--proton collisions. 
The main track-based alignment method uses the Millepede global approach.
An iterative local method was developed and used as well.
We present the results obtained for the ITS
alignment using about $10^5$ charged tracks from cosmic rays that have 
been collected during summer 2008, with the ALICE solenoidal magnet
switched off.}


\begin{document}


\addcontentsline{toc}{section}{ALICE collaboration}

\section*{ALICE collaboration}

\author{
K.~Aamodt\inst{78},
N.~Abel\inst{43},
U.~Abeysekara\inst{30},
A.~Abrahantes~Quintana\inst{42},
D.~Adamov\'{a}\inst{86},
M.M.~Aggarwal\inst{25},
G.~Aglieri~Rinella\inst{40},
A.G.~Agocs\inst{18},
S.~Aguilar~Salazar\inst{66},
Z.~Ahammed\inst{55}
% ...
\renewcommand{\notesname}{Affiliation notes}
\endnotetext[1]{Also at\inst{26}}
\endnotetext[2]{Also at\inst{36}}
\endnotetext[3]{Now at\inst{80}}
\endnotetext[4]{Now at\inst{32}}
\endnotetext[5]{Now at\inst{37}}
\endnotetext[6]{Now at\inst{22}}
\endnotetext[7]{Now at\inst{17}}
\endnotetext[8]{Also at\inst{46}}
\endnotetext[9]{Now at\inst{57}}
\endnotetext[10]{Now at\inst{40}}
\endnotetext[11]{Now at\inst{49}}
\endnotetext[12]{Now at\inst{99}}
\endnotetext[13]{Now at\inst{6}}
\endnotetext[14]{Now at\inst{72}}
\endnotetext[15]{Now at: University of Technology and Austrian Academy of Sciences, Vienna, Austria}
\endnotetext[16]{Also at\inst{60}}
\endnotetext[17]{Also at\inst{40}}
\endnotetext[18]{Now at \inst{115}}
\endnotetext[19]{Deceased}
\endnotetext[20]{Now at\inst{74}}
\endnotetext[21]{Now at\inst{105}}
\endnotetext[22]{Also at \inst{114}}
\endnotetext[23]{Now at\inst{5}}
\endnotetext[24]{Also at\inst{41}}
\endnotetext[25]{Now at\inst{101}}
\endnotetext[26]{Now at\inst{30}}
\endnotetext[27]{Now at\inst{89}}
\endnotetext[28]{Also at\inst{78}}
\endnotetext[29]{Now at\inst{44}}
\endnotetext[30]{Now at\inst{33}}
\endnotetext[31]{Now at\inst{8}}
\endnotetext[32]{Now at\inst{4}}
\endnotetext[33]{Also at\inst{72}}
\endnotetext[34]{Now at\inst{73}}
\endnotetext[35]{Now at\inst{62}}
\endnotetext[36]{Now at: Centre de Calcul IN2P3, Lyon, France}
\endnotetext[37]{Now at\inst{55}}
\endnotetext[38]{Also at\inst{79}}
\endnotetext[39]{Also at\inst{14}}
\endnotetext[40]{Also at Dipartimento di Fisica dell\'{ }Universit\`{a}, Udine, Italy}
\bigskip
\theendnotes
\section*{Collaboration institutes}

{\footnotesize
\inst{1}
Dipartimento di Scienze e Tecnologie Avanzate dell'Universit\`{a} del Piemonte Orientale and Gruppo Collegato INFN, Alessandria, Italy
\\\noindent\inst{2}
Department of Physics Aligarh Muslim University, Aligarh, India
\\\noindent\inst{3}
National Institute for Nuclear and High Energy Physics (NIKHEF), Amsterdam, Netherlands \\\noindent\inst{4}
Physics Department, University of Athens, Athens, Greece
\\\noindent\inst{5}
Dipartimento Interateneo di Fisica `M.~Merlin' and Sezione INFN, Bari, Italy
\\\noindent\inst{6}
Sezione INFN, Bari, Italy
\\\noindent\inst{7}
China Institute of Atomic Energy, Beijing, China
\\\noindent\inst{8}
Department of Physics and Technology, University of Bergen, Bergen, Norway
\\\noindent\inst{9}
Faculty of Engineering, Bergen University College, Bergen, Norway
\\\noindent\inst{10}
Lawrence Berkeley National Laboratory, Berkeley, California, United States
\\\noindent\inst{11}
Institute of Physics, Bhubaneswar, India
\\\noindent\inst{12}
School of Physics and Astronomy, University of Birmingham, Birmingham, United Kingdom
\\\noindent\inst{13}
Dipartimento di Fisica dell'Universit\`{a} and Sezione INFN, Bologna, Italy
\\\noindent\inst{14}
Sezione INFN, Bologna, Italy
\\\noindent\inst{15}
Faculty of Mathematics, Physics and Informatics, Comenius University, Bratislava, Slovakia
\\\noindent\inst{16}
Institute of Space Sciences (ISS), Bucharest, Romania
\\\noindent\inst{17}
National Institute for Physics and Nuclear Engineering, Bucharest, Romania
\\\noindent\inst{18}
KFKI Research Institute for Particle and Nuclear Physics, Hungarian Academy of Sciences, Budapest, Hungary
\\\noindent\inst{19}
Dipartimento di Fisica dell'Universit\`{a} and Sezione INFN, Cagliari, Italy
\\\noindent\inst{20}
Sezione INFN, Cagliari, Italy
\\\noindent\inst{21}
Universidade Estadual de Campinas (UNICAMP), Campinas, Brazil
\\\noindent\inst{22}
Physics Department, University of Cape Town, iThemba Laboratories, Cape Town, South Africa
\\\noindent\inst{23}
Dipartimento di Fisica e Astronomia dell'Universit\`{a} and Sezione INFN, Catania, Italy
\\\noindent\inst{24}
Sezione INFN, Catania, Italy
\\\noindent\inst{25}
Physics Department, Panjab University, Chandigarh, India
\\\noindent\inst{26}
Laboratoire de Physique Corpusculaire (LPC), Clermont Universit\'{e}, Universit\'{e} Blaise Pascal, CNRS--IN2P3, Clermont-Ferrand, France
\\\noindent\inst{27}
Department of Physics, Ohio State University, Columbus, Ohio, United States
\\\noindent\inst{28}
Niels Bohr Institute, University of Copenhagen, Copenhagen, Denmark
\\\noindent\inst{29}
The Henryk Niewodniczanski Institute of Nuclear Physics, Polish Academy of Sciences, Cracow, Poland
\\\noindent\inst{30}
Physics Department, Creighton University, Omaha, Nebraska, United States
\\\noindent\inst{31}
Universidad Aut\'{o}noma de Sinaloa, Culiac\'{a}n, Mexico
\\\noindent\inst{32}
ExtreMe Matter Institute EMMI, GSI Helmholtzzentrum f\"{u}r Schwerionenforschung, Darmstadt, Germany
\\\noindent\inst{33}
Institut f\"{u}r Kernphysik, Technische Universit\"{a}t Darmstadt, Darmstadt, Germany
\\\noindent\inst{34}
Wayne State University, Detroit, Michigan, United States
\\\noindent\inst{35}
Joint Institute for Nuclear Research (JINR), Dubna, Russia
\\\noindent\inst{36}
Frankfurt Institute for Advanced Studies, Johann Wolfgang Goethe-Universit\"{a}t Frankfurt, Frankfurt, Germany
\\\noindent\inst{37}
Institut f\"{u}r Kernphysik, Johann Wolfgang Goethe-Universit\"{a}t Frankfurt, Frankfurt, Germany
\\\noindent\inst{38}
Laboratori Nazionali di Frascati, INFN, Frascati, Italy
\\\noindent\inst{39}
Petersburg Nuclear Physics Institute, Gatchina, Russia
\\\noindent\inst{40}
European Organization for Nuclear Research (CERN), Geneva, Switzerland
\\\noindent\inst{41}
Laboratoire de Physique Subatomique et de Cosmologie (LPSC), Universit\'{e} Joseph Fourier, CNRS-IN2P3, Institut Polytechnique de Grenoble, Grenoble, France
\\\noindent\inst{42}
Centro de Aplicaciones Tecnol\'{o}gicas y Desarrollo Nuclear (CEADEN), Havana, Cuba
\\\noindent\inst{43}
Kirchhoff-Institut f\"{u}r Physik, Ruprecht-Karls-Universit\"{a}t Heidelberg, Heidelberg, Germany
\\\noindent\inst{44}
Physikalisches Institut, Ruprecht-Karls-Universit\"{a}t Heidelberg, Heidelberg, Germany
\\\noindent\inst{45}
Hiroshima University, Hiroshima, Japan
\\\noindent\inst{46}
University of Houston, Houston, Texas, United States
\\\noindent\inst{47}
Physics Department, University of Rajasthan, Jaipur, India
\\\noindent\inst{48}
Physics Department, University of Jammu, Jammu, India
\\\noindent\inst{49}
Helsinki Institute of Physics (HIP) and University of Jyv\"{a}skyl\"{a}, Jyv\"{a}skyl\"{a}, Finland
\\\noindent\inst{50}
Kangnung National University, Kangnung, South Korea
\\\noindent\inst{51}
Scientific Research Technological Institute of Instrument Engineering, Kharkov, Ukraine
\\\noindent\inst{52}
Bogolyubov Institute for Theoretical Physics, Kiev, Ukraine
\\\noindent\inst{53}
University of Tennessee, Knoxville, Tennessee, United States
\\\noindent\inst{54}
Saha Institute of Nuclear Physics, Kolkata, India
\\\noindent\inst{55}
Variable Energy Cyclotron Centre, Kolkata, India
\\\noindent\inst{56}
Fachhochschule K\"{o}ln, K\"{o}ln, Germany
\\\noindent\inst{57}
Faculty of Science, P.J.~\v{S}af\'{a}rik University, Ko\v{s}ice, Slovakia
\\\noindent\inst{58}
Institute of Experimental Physics, Slovak Academy of Sciences, Ko\v{s}ice, Slovakia
\\\noindent\inst{59}
Laboratori Nazionali di Legnaro, INFN, Legnaro, Italy
\\\noindent\inst{60}
Lawrence Livermore National Laboratory, Livermore, California, United States
\\\noindent\inst{61}
Division of Experimental High Energy Physics, University of Lund, Lund, Sweden
\\\noindent\inst{62}
Universit\'{e} de Lyon 1, CNRS/IN2P3, Institut de Physique Nucl\'{e}aire de Lyon, Lyon, France
\\\noindent\inst{63}
Centro de Investigaciones Energ\'{e}ticas Medioambientales y Tecnol\'{o}gicas (CIEMAT), Madrid, Spain
\\\noindent\inst{64}
Centro de Investigaci\'{o}n y de Estudios Avanzados (CINVESTAV), Mexico City and M\'{e}rida, Mexico
\\\noindent\inst{65}
Instituto de Ciencias Nucleares, Universidad Nacional Aut\'{o}noma de M\'{e}xico, Mexico City, Mexico
\\\noindent\inst{66}
Instituto de F\'{\i}sica, Universidad Nacional Aut\'{o}noma de M\'{e}xico, Mexico City, Mexico
\\\noindent\inst{67}
Institute for Nuclear Research, Academy of Sciences, Moscow, Russia
\\\noindent\inst{68}
Institute for Theoretical and Experimental Physics, Moscow, Russia
\\\noindent\inst{69}
Moscow Engineering Physics Institute, Moscow, Russia
\\\noindent\inst{70}
Russian Research Centre Kurchatov Institute, Moscow, Russia
\\\noindent\inst{71}
Indian Institute of Technology, Mumbai, India
\\\noindent\inst{72}
Institut f\"{u}r Kernphysik, Westf\"{a}lische Wilhelms-Universit\"{a}t M\"{u}nster, M\"{u}nster, Germany
\\\noindent\inst{73}
SUBATECH, Ecole des Mines de Nantes, Universit\'{e} de Nantes, CNRS-IN2P3, Nantes, France
\\\noindent\inst{74}
Yale University, New Haven, Connecticut, United States
\\\noindent\inst{75}
Budker Institute for Nuclear Physics, Novosibirsk, Russia
\\\noindent\inst{76}
Oak Ridge National Laboratory, Oak Ridge, Tennessee, United States
\\\noindent\inst{77}
Institut de Physique Nucl\'{e}aire d'Orsay (IPNO), Universit\'{e} Paris-Sud, CNRS-IN2P3, Orsay, France
\\\noindent\inst{78}
Department of Physics, University of Oslo, Oslo, Norway
\\\noindent\inst{79}
Dipartimento di Fisica dell'Universit\`{a} and Sezione INFN, Padova, Italy
\\\noindent\inst{80}
Sezione INFN, Padova, Italy
\\\noindent\inst{81}
Faculty of Nuclear Sciences and Physical Engineering, Czech Technical University in Prague, Prague, Czech Republic
\\\noindent\inst{82}
Institute of Physics, Academy of Sciences of the Czech Republic, Prague, Czech Republic
\\\noindent\inst{83}
Institute for High Energy Physics, Protvino, Russia
\\\noindent\inst{84}
Benem\'{e}rita Universidad Aut\'{o}noma de Puebla, Puebla, Mexico
\\\noindent\inst{85}
Pusan National University, Pusan, South Korea
\\\noindent\inst{86}
Nuclear Physics Institute, Academy of Sciences of the Czech Republic, \v{R}e\v{z} u Prahy, Czech Republic
\\\noindent\inst{87}
Dipartimento di Fisica dell'Universit\`{a} `La Sapienza' and Sezione INFN, Rome, Italy
\\\noindent\inst{88}
Sezione INFN, Rome, Italy
\\\noindent\inst{89}
Commissariat \`{a} l'Energie Atomique, IRFU, Saclay, France
\\\noindent\inst{90}
Dipartimento di Fisica `E.R.~Caianiello' dell'Universit\`{a} and Sezione INFN, Salerno, Italy
\\\noindent\inst{91}
California Polytechnic State University, San Luis Obispo, California, United States
\\\noindent\inst{92}
Departamento de F\'{\i}sica de Part\'{\i}culas and IGFAE, Universidad de Santiago de Compostela, Santiago de Compostela, Spain
\\\noindent\inst{93}
Universidade de S\~{a}o Paulo (USP), S\~{a}o Paulo, Brazil
\\\noindent\inst{94}
Russian Federal Nuclear Center (VNIIEF), Sarov, Russia
\\\noindent\inst{95}
Department of Physics, Sejong University, Seoul, South Korea
\\\noindent\inst{96}
Yonsei University, Seoul, South Korea
\\\noindent\inst{97}
Technical University of Split FESB, Split, Croatia
\\\noindent\inst{98}
V.~Fock Institute for Physics, St. Petersburg State University, St. Petersburg, Russia
\\\noindent\inst{99}
Institut Pluridisciplinaire Hubert Curien (IPHC), Universit\'{e} de Strasbourg, CNRS-IN2P3, Strasbourg, France
\\\noindent\inst{100}
University of Tokyo, Tokyo, Japan
\\\noindent\inst{101}
Dipartimento di Fisica Sperimentale dell'Universit\`{a} and Sezione INFN, Turin, Italy \\\noindent\inst{102}
Sezione INFN, Turin, Italy
\\\noindent\inst{103}
Dipartimento di Fisica dell'Universit\`{a} and Sezione INFN, Trieste, Italy
\\\noindent\inst{104}
Sezione INFN, Trieste, Italy
\\\noindent\inst{105}
University of Tsukuba, Tsukuba, Japan
\\\noindent\inst{106}
Institute for Subatomic Physics, Utrecht University, Utrecht, Netherlands
\\\noindent\inst{107}
Soltan Institute for Nuclear Studies, Warsaw, Poland
\\\noindent\inst{108}
Warsaw University of Technology, Warsaw, Poland
\\\noindent\inst{109}
Purdue University, West Lafayette, Indiana, United States
\\\noindent\inst{110}
Zentrum f\"{u}r Technologietransfer und Telekommunikation (ZTT), Fachhochschule Worms, Worms, Germany
\\\noindent\inst{111}
Hua-Zhong Normal University, Wuhan, China
\\\noindent\inst{112}
Yerevan Physics Institute, Yerevan, Armenia
\\\noindent\inst{113}     
Rudjer Bo\v{s}kovi\'{c} Institute, Zagreb, Croatia
\\\noindent\inst{114}     
Centro Fermi -- Centro Studi e Ricerche e Museo Storico della Fisica ``Enrico Fermi'', Rome, Italy
\\\noindent\inst{115}
Secci\'{o}n F\'{\i}sica, Departamento de Ciencias, Pontificia Universidad Cat\'{o}lica del Per\'{u}, Lima, Peru
}

\vspace{1cm}
\noindent
Corresponding author: Andrea Dainese (\texttt{andrea.dainese@pd.infn.it})
}

\clearpage

\section{Introduction}
\label{sec:intro}


\noindent
The ALICE experiment~\cite{aliceJINST} will study nucleus--nucleus, 
proton--proton and proton--nucleus 
collisions at the CERN Large Hadron Collider (LHC).
The main physics goal of the experiment is to investigate 
the properties of strongly-interacting matter
in the conditions of high energy 
density ($>10~\gev/\fm^3$) and high temperature ($\gsim 0.2~\gev$),
expected to be reached in central \mbox{Pb--Pb} collisions at 
$\sqrtsNN=5.5~\tev$. 
Under these 
conditions, according to lattice QCD calculations, quark confinement into 
colourless hadrons should be removed and
a deconfined Quark--Gluon Plasma should be formed~\cite{PPR1}.
In the past two decades, experiments at CERN-SPS ($\sqrtsNN=17.3~\gev$) 
and BNL-RHIC ($\sqrtsNN=200~\gev$) have gathered
ample evidence for the formation of this state of matter~\cite{QGPexp}.

\begin{figure}[!b]
\begin{center}
\includegraphics[width=12cm]{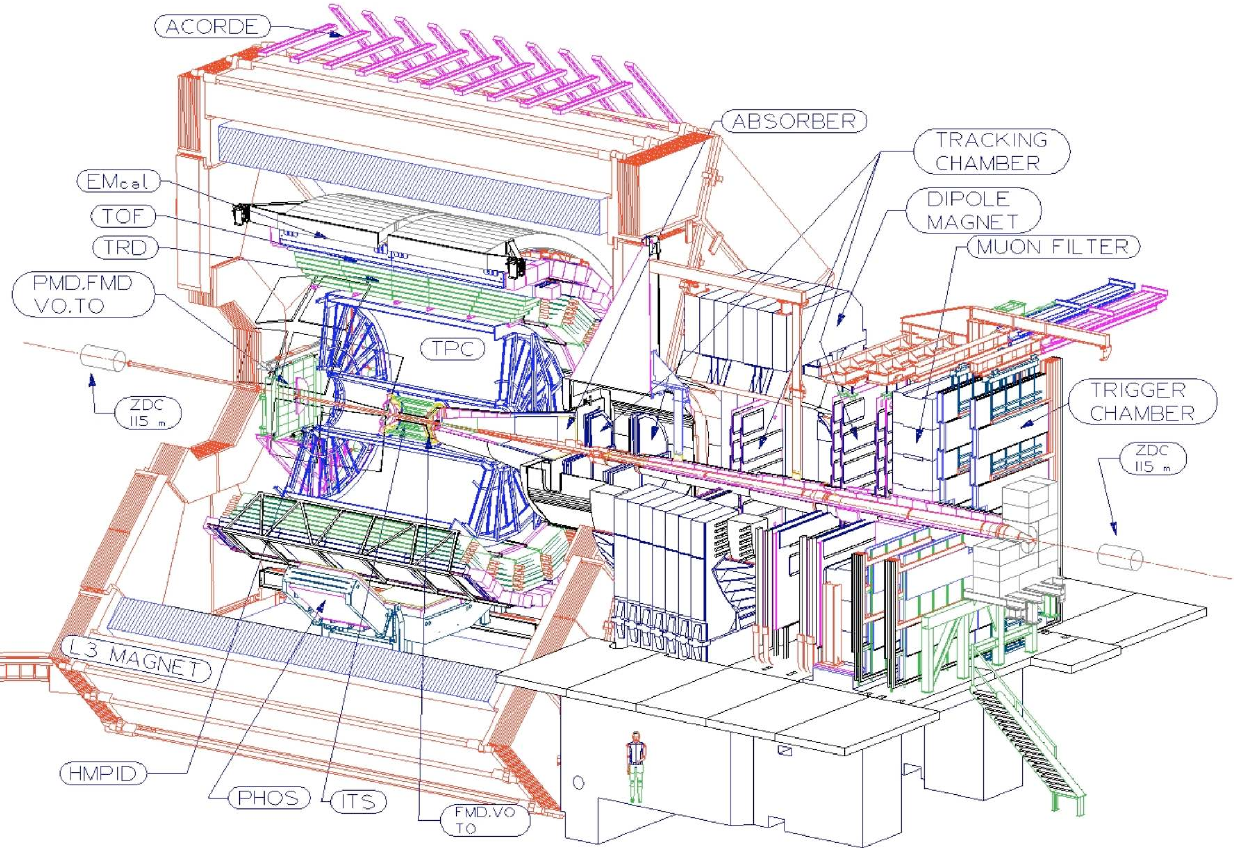}
\caption{General layout of the ALICE experiment~\cite{aliceJINST}.}
\label{fig:alice}
\end{center}
\end{figure}

The ALICE experimental apparatus, shown in Fig.~\ref{fig:alice},
 consists of a central barrel, a forward muon spectrometer and a set of small detectors in the forward 
regions for trigger and other functions.
The coverage of the central barrel detectors
allows the tracking of particles emitted within a pseudo-rapidity range
$|\eta| < 0.9$ over the full azimuth. The central barrel is surrounded by
the large L3 magnet that provides a field
${\rm B}=0.5~\mathrm{T}$.

The ITS (Inner Tracking System) is a cylindrically-shaped silicon 
tracker that surrounds the interaction region.
It consists of six layers,
with radii between 3.9~cm and 43.0~cm, 
covering the pseudo-rapidity range $|\eta|<0.9$.
The two innermost layers are equipped with Silicon Pixel Detectors (SPD), 
the two intermediate layers contain Silicon Drift Detectors (SDD), while
Silicon Strip Detectors (SSD) are used on the two outermost layers.
The main task of the ITS is to provide precise track and vertex reconstruction
close to the interaction point.
In particular, 
the ITS was designed with the aim to improve the position, angle, and momentum resolution for 
tracks reconstructed in the Time Projection Chamber (TPC), 
to identify the secondary vertices from the decay of
hyperons and heavy flavoured hadrons,
to reconstruct the interaction vertex 
with a resolution better
than 100~$\mu$m, and 
to recover 
particles that are missed by the TPC due to acceptance limitations 
(very low momentum particles not reaching the TPC 
and very high momentum ones propagating along the 10\% inactive area 
between adjacent TPC chambers).

The measurement of charm and beauty hadron production in 
Pb--Pb collisions at the LHC is one of the main items of the ALICE 
physics program, because it will allow to 
investigate the mechanisms of heavy-quark propagation and
 hadronization in the large, hot and dense medium
formed in high-energy heavy-ion collisions and it will serve as a 
reference for the study of the effects of the medium 
on quarkonia states~\cite{PPR2}. 
To measure the separation, 
from the interaction vertex, of the decay vertices of heavy flavoured hadrons,
which have mean proper decay lengths $c\tau\sim 100$--$500~\mum$,
requires a resolution on the track impact parameter 
(distance of closest approach to the vertex) well below $100~\mum$. 
This requirement is met by the ITS. 
The design position resolution
in the plane transverse to the beam line 
for charged-pion 
tracks reconstructed in the TPC and in the ITS
is expected to be approximately  
$10~\mum+53~\mum/(\pt\sqrt{\sin\theta})$, 
where $\pt$ is the transverse momentum in $\gev/c$ and $\theta$ is the polar angle with respect to the beam line~\cite{PPR2}. 
The ITS is made of thousands of separate modules, 
whose position is different from the ideal due to the limitations associated
with the assembly and integration of the different components, and the forces
these components experience.
In order to achieve the required high precision on the track parameters, the
relative position (location and orientation) of every module needs to be 
determined precisely. 
We refer to the procedure used to determine the modules
relative position as alignment.
The ITS alignment procedure starts from the positioning survey
measurements performed during the assembly, and is refined using tracks 
from cosmic-ray muons and from particles produced in LHC pp collisions. 
Two independent methods, based on tracks-to-measured-points residuals 
minimization, are considered. 
The first method uses the Millepede approach~\cite{refMP1}, where a 
global fit to all residuals is performed, extracting all the alignment
 parameters simultaneously. The second method performs a (local) minimization for each single module 
and accounts for correlations between modules by iterating the procedure until convergence is reached.

In this article, we present the alignment methods for the ITS and
the results obtained using the cosmic-data sample collected during summer
2008 with $\rm B=0$ (a small data set with $\rm B=\pm 0.5~T$ was also collected; 
we used it for a few specific validation checks).
In section~\ref{sec:ITSdescription} we describe in detail the ITS detector layout 
and in section~\ref{sec:strategy} we discuss the strategy adopted for the alignment. In section~\ref{sec:cosmics}
we describe the 2008 sample of cosmic-muon data.
These data were used to validate the available survey measurements 
(section~\ref{sec:survey}) and to apply the track-based alignment algorithms:
the Millepede method (section~\ref{sec:millepede}) and a local 
method that we are developing (section~\ref{sec:iterative}).
We draw conclusions in section~\ref{sec:conclusions}.


\section{ITS detector layout}
\label{sec:ITSdescription}


\begin{figure}[!t]
\centering
\resizebox{0.67\textwidth}{!}{%
\includegraphics*[]{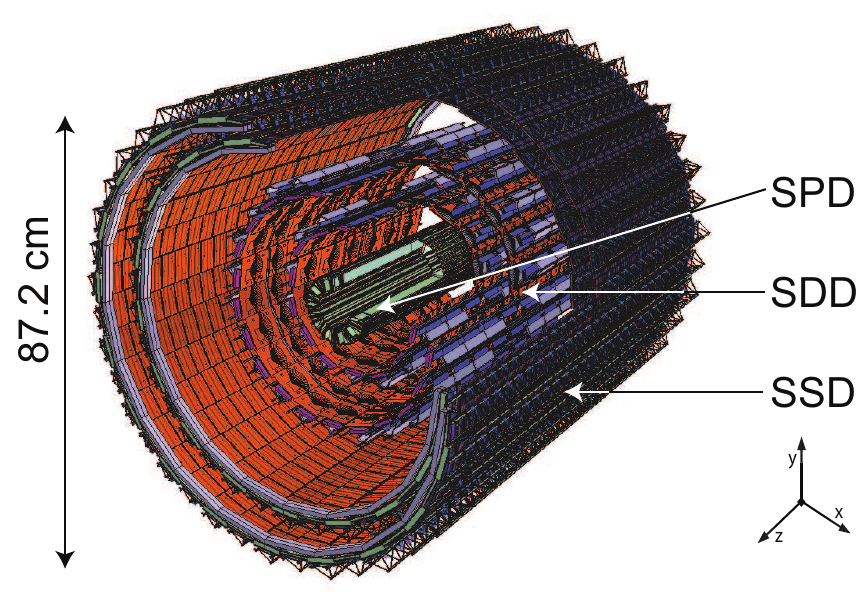}}
\hfill
\resizebox{0.25\textwidth}{!}{%
\includegraphics*[]{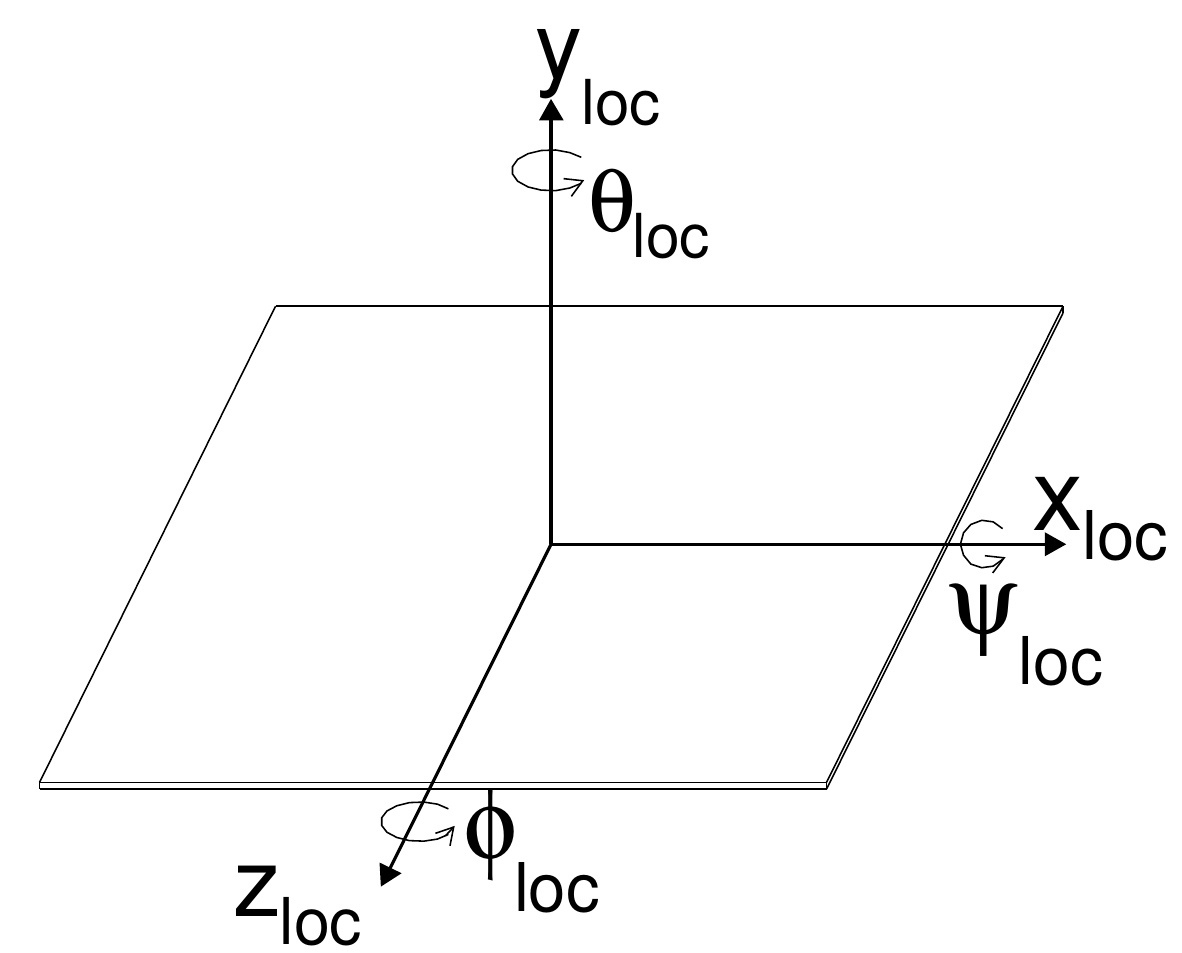}}
\caption{Layout of the ITS (left) and orientation of the ALICE global (middle) 
and ITS-module local (right) reference systems. The global reference 
system has indeed its origin in the middle of the ITS, so that the $z$ direction
coincides with the beam line.}
\label{fig:itsscheme}
\end{figure}

\noindent
The geometrical layout of the ITS layers
is shown in the left-hand panel of Fig.~\ref{fig:itsscheme},
 as it is implemented in the 
ALICE simulation and reconstruction software framework (AliRoot~\cite{aliroot}).
The ALICE global reference system has 
the $z$ axis on the beam line, the $x$ axis in the LHC (horizontal) plane,
pointing to the centre of the accelerator, and the $y$ axis pointing upward.
The axis of the ITS barrel coincides with the $z$ axis.
The module local reference system (Fig.~\ref{fig:itsscheme}, right)
is defined with the $x_{\rm loc}$ and $z_{\rm loc}$ axes on 
the sensor plane and with the $z_{\rm loc}$ axis in the same direction as the
global $z$ axis. The local $x$ direction is approximately equivalent to 
the global $r\varphi$.
The alignment degrees of freedom of the module are translations in 
$x_{\rm loc}$, $y_{\rm loc}$, $z_{\rm loc}$, and rotations by angles $\psi_{\rm loc}$, $\theta_{\rm loc}$, $\varphi_{\rm loc}$,
about the $x_{\rm loc}$, $y_{\rm loc}$, $z_{\rm loc}$ axes, respectively\footnote{The alignment 
transformation can be expressed equivalently in terms of the local 
or global coordinates.}.

The ITS geometry in AliRoot is described in full detail, down to the level of 
all mechanical structures and single electronic components,
using the ROOT~\cite{root} geometrical modeler. 
This detailed geometry is
used in Monte Carlo simulations and in the track reconstruction procedures,
thereby accounting for the exact position of the sensor modules and 
of all the passive material that determine particle scattering
and energy loss.

The geometrical parameters of the layers (radial position, length along
beam axis, number of modules, spatial resolution, and material budget) 
are summarized in Table~\ref{tab:ITSlayers}.
The material budget reported in the table takes into account the 
$\phi$-averaged material (including the sensors, electronics, cabling, support structures, 
and cooling) associated with radial paths through each layer.
Another 1.30\% of radiation length comes from the thermal
shields and supports installed between SPD and SDD barrels and between SDD
and SSD barrels, thus making the total material budget for perpendicular 
tracks equal to 7.66\% of $X_0$.

\begin{table}[t!]
\caption{Characteristics of the six ITS layers.} 
\label{tab:ITSlayers}
\centering
\begin{tabular}{|c|l|c|c|c|c|c|c|} 
\hline 
 & & & & Number &
Active Area & & Material\\
Layer  & Type     &  $r$ [cm]  & $\pm z$ [cm]              & of   &
per module  & Resolution & budget\\
      &      &          &              &    modules &
  $r\varphi$~$\times$~$z$ [mm$^2$] & $r\varphi$~$\times$~$z$ [$\mum^2$] & $X/X_0$ [\%] \\
\hline
\hline
 1  & pixel & \phantom{0}3.9 & 14.1  & \phantom{0}80 & 
12.8$\times$70.7 & \phantom{0}12$\times$100 & 1.14\\
 2  & pixel & \phantom{0}7.6 & 14.1  & 160           &
12.8$\times$70.7 & \phantom{0}12$\times$100  & 1.14\\
 3  & drift & 15.0           & 22.2  & \phantom{0}84  &
70.17$\times$75.26 & \phantom{0}35$\times$25\phantom{0} & 1.13\\
 4  & drift & 23.9           & 29.7  & 176            &
70.17$\times$75.26 & \phantom{0}35$\times$25\phantom{0} & 1.26\\
 5  & strip & 38.0           & 43.1  & 748    &
73$\times$40 & \phantom{0}20$\times$830        & 0.83 \\
 6  & strip & 43.0           & 48.9  & 950         &
73$\times$40 & \phantom{0}20$\times$830   & 0.86 \\
\hline
\end{tabular}
\end{table}

In the following, the features of each of the
three sub-detectors (SPD, SDD and SSD) that
are relevant for alignment issues are described 
(for more details see~\cite{aliceJINST}). 

\subsection{Silicon Pixel Detector (SPD)}
\label{sec:SPDdescription}

\begin{figure}[!b]
\centering
\begin{minipage}{0.49\textwidth}
\centering
\resizebox{0.9\textwidth}{!}{
\includegraphics*{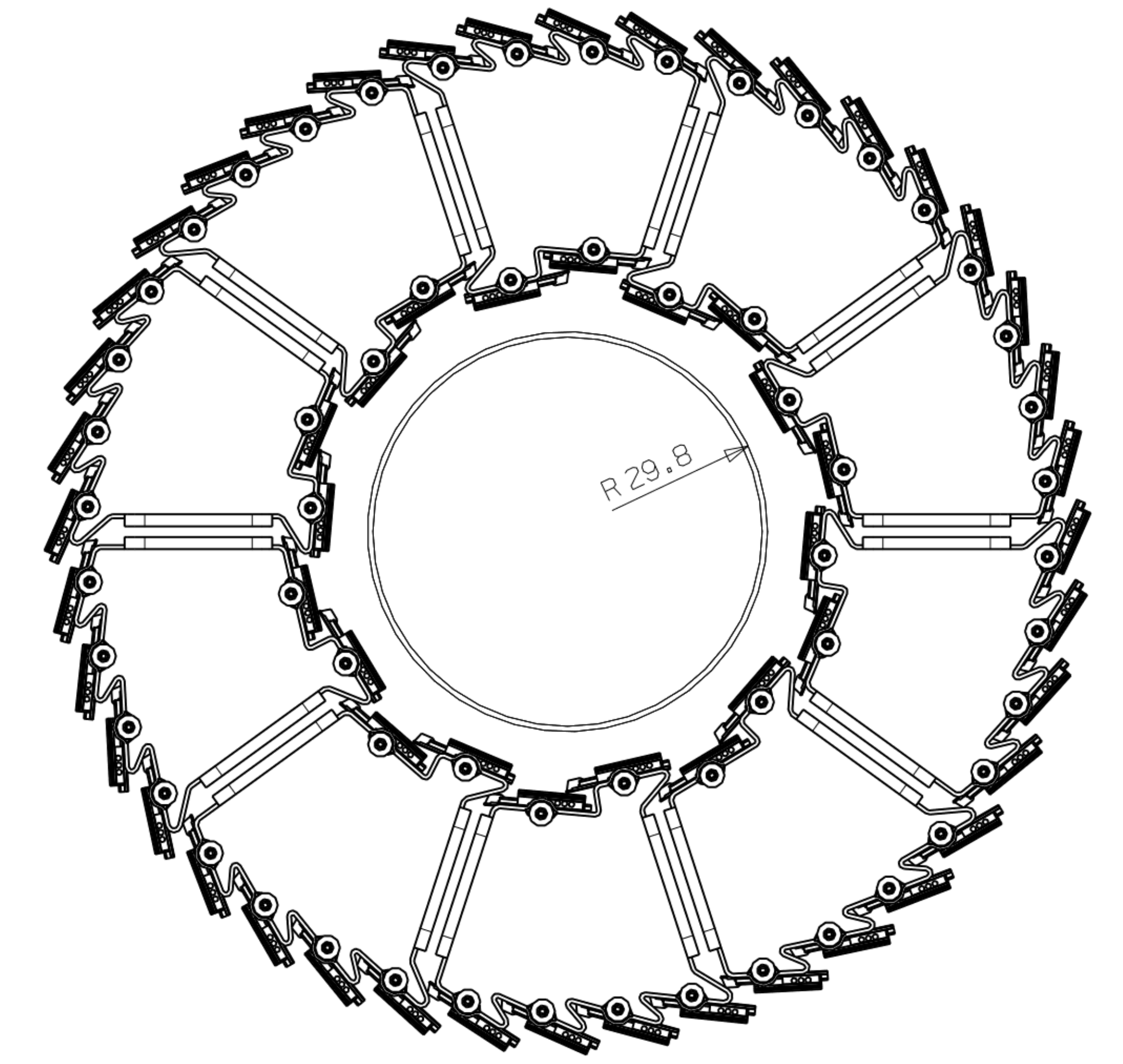}}
\end{minipage}\hfill
\begin{minipage}{0.49\textwidth}
\resizebox{0.9\textwidth}{!}{
\rotatebox{270}{%
\includegraphics*[angle=+90]{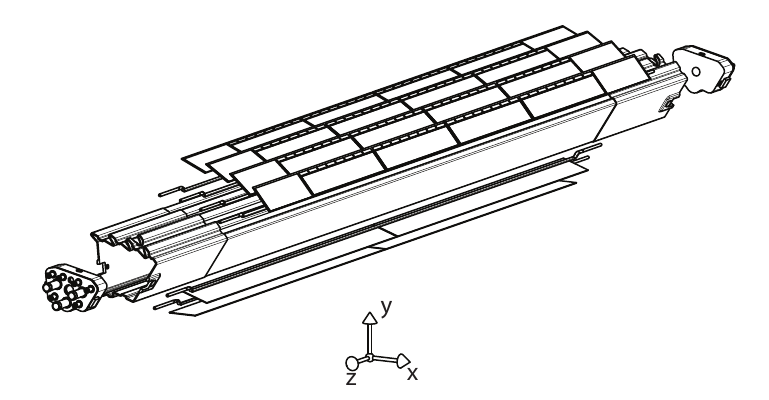}}}
\end{minipage}
\caption{SPD drawings. Left: the SPD barrel and the beam pipe (radius in mm). Right: a Carbon Fibre Support Sector. }
\label{fig:spdsector}
\end{figure}

\noindent
The basic building block of the ALICE SPD is a module consisting of a
two-dimensional sensor matrix of reverse-biased 
silicon detector diodes bump-bonded to 5 front-end chips.
The sensor matrix consists of 256 $\times$ 160 cells,
each measuring 50~$\mu$m ($r\varphi$) by 425~$\mu$m ($z$).

Two modules are mounted together along the $z$ direction to form a 141.6~mm
long half-stave.
Two mirrored half-staves are attached, head-to-head along the $z$ direction,
 to a 
carbon-fibre support sector, which also provides cooling.
Each sector (see Fig.~\ref{fig:spdsector}, right) supports six staves: 
two on the inner layer and four on the 
outer layer.
The sensors are mounted in such a way that there is a 2\% overlap between
the active regions in $r\varphi$, but along z there is a gap between each two 
consecutive sensors.
Five sectors are then mounted together to form a half-barrel and
finally the two (top and bottom) half-barrels are mounted around the beam pipe 
to close the full barrel (shown in the left-hand side of 
Fig.~\ref{fig:spdsector}), which is actually composed of 10 sectors. 
In total, the SPD includes 60 staves, consisting of 240 modules with 
1200 readout chips for a total of $9.8 \times 10^{6}$ cells.

The spatial precision of the SPD sensor is determined by the pixel cell size,
the track incidence angle on the detector, and by the 
threshold applied in the readout electronics.
The values of resolution along $r\varphi$ and $z$ extracted from beam tests
are 12 and 100~$\mu$m, respectively.

\subsection{Silicon Drift Detector (SDD)}
\label{sec:SDDdescription}

\begin{figure}[!b]
\includegraphics[width=0.50\textwidth]{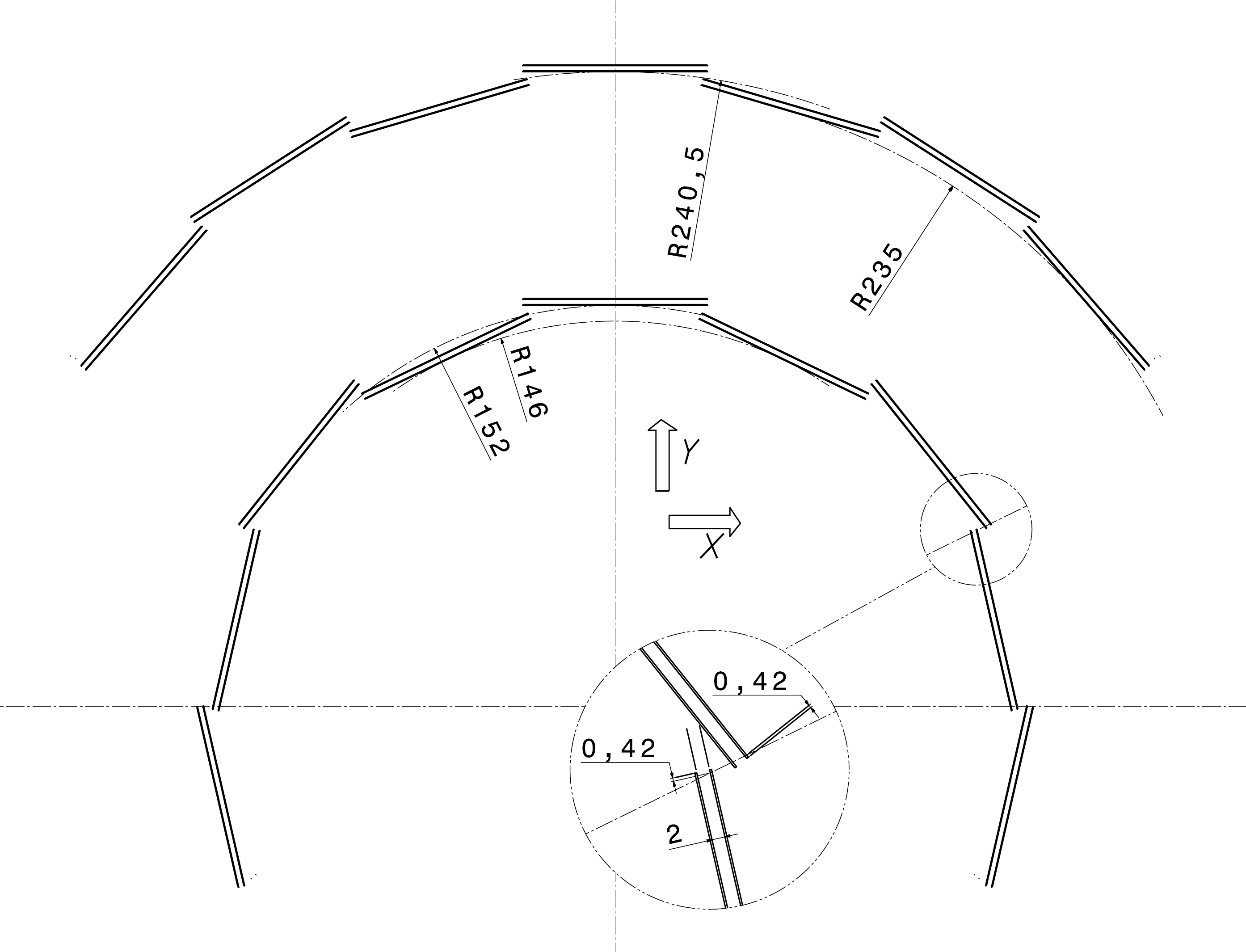}
\hfill
\includegraphics[width=0.40\textwidth]{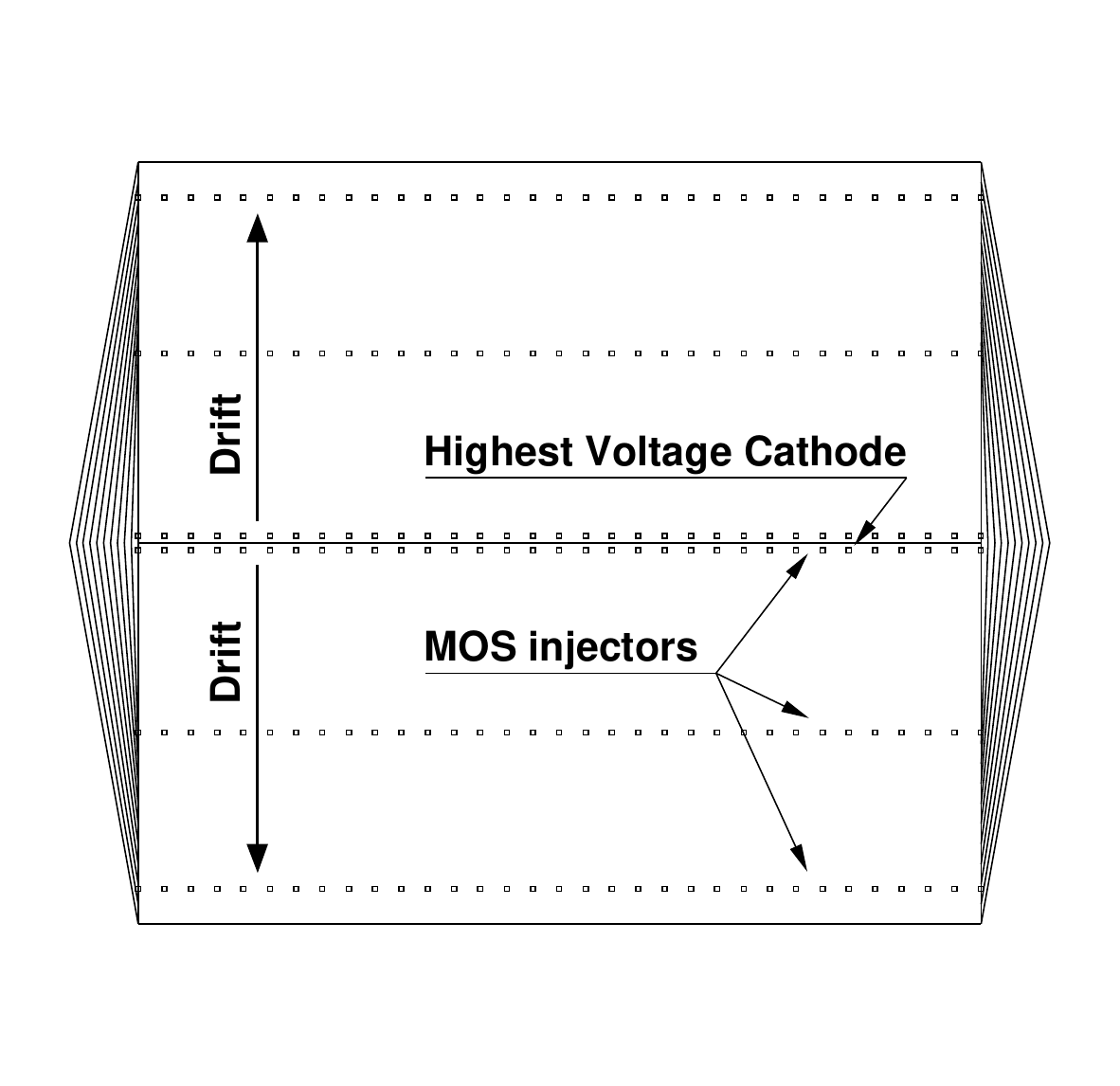}
\caption{Left: scheme of the SDD layers. 
Right: scheme of a SDD module, where the drift direction is parallel to the 
$x_{\rm loc}$ coordinate.
Units are millimeters.}
\label{fig:sddstruct}
\end{figure}

\noindent
The basic building block of the ALICE SDD~\cite{SDDdescr} is a module 
divided into two drift regions
where electrons move in opposite directions under a drift field of
$\approx500$~V/cm (see Fig.~\ref{fig:sddstruct}, right), with 
hybrids housing the front-end electronics on either side.
The SDD modules are mounted on linear structures called ladders. 
There are 14 ladders with six modules each on the inner SDD layer (layer 3), 
and 22 ladders with eight modules each on the outer SDD layer (layer 4). 
Modules and  ladders are assembled to
have an overlap of the sensitive areas larger than 580~$\mu$m in both
$r\varphi$ and $z$ directions, so as to provide full angular coverage
over the pseudo-rapidity range $|\eta|<0.9$
(Fig.~\ref{fig:sddstruct}, left).

The modules are attached to the ladder space frame
and have their anode rows parallel to the ladder axis ($z$). 
The ladders are mounted on a support structure made of two cones and 
four support rings to form the two cylindrical layers~\cite{giraudo}. 
The support rings are mechanically fixed to the cones and bear 
ruby spheres, used as a reference for the ladder positioning as well as
for the geometrical survey of the module positions in the ladder
reference system.

The $z$ coordinate is 
reconstructed from the centroid of the collected charge along the anodes.
The position along the drift coordinate ($x_{\rm loc}\approx r\varphi$) 
is reconstructed
starting from the measured drift time with respect to the trigger time.
An unbiased reconstruction of the $x_{\rm loc}$ coordinate requires therefore to
know with good precision the drift velocity and the time-zero ($t_0$), 
which is the measured drift time for particles with zero drift distance.
The drift velocity depends on temperature (as $T^{-2.4}$) and it is 
therefore sensitive to temperature gradients in the SDD volume and
to temperature variations with time. 
Hence, it is important to 
calibrate this parameter frequently during the data taking.
Three rows of 33 MOS charge injectors are implanted at known
distances from the collection anodes
in each of the two drift regions of a SDD module~\cite{SDDinj}
for this purpose, as sketched
in Fig.~\ref{fig:sddstruct} (right).
Finally, a correction for non-uniformity
of the drift field (due to non-linearities in the voltage
divider and, for a few modules, also due to significant inhomogeneities in 
dopant concentration) has to  be applied. This correction 
is extracted from measurements of the systematic deviations 
between charge injection position and reconstructed coordinates that was 
performed on all the 260 SDD modules with an infrared laser~\cite{SDDmaps}.

The spatial precision of the SDD detectors, as 
obtained during beam tests of full-size prototypes, is 
on average 35~$\mu$m along the drift direction $x_{\rm loc}$ and 25~$\mu$m
for the anode coordinate $z_{\rm loc}$.

\subsection{Silicon Strip Detector (SSD)}
\label{sec:SSDdescription}

\begin{figure}[!b]
\centering
\resizebox{0.9\textwidth}{!}{
\includegraphics*[]{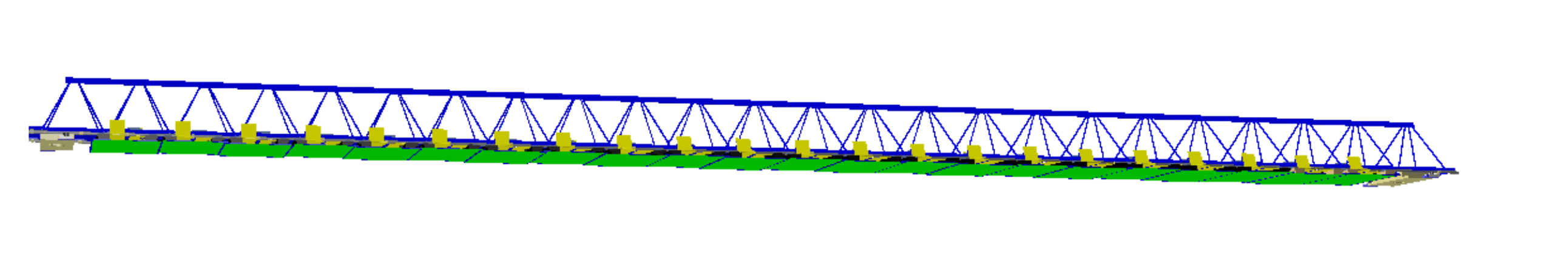}}
\caption{View of one SSD ladder (from layer 5)
as described in the AliRoot geometry.}
\label{fig:ssdladder}
\end{figure}

The basic building block of the ALICE SSD
is a module composed of one double-sided strip
detector connected to two hybrids hosting the front-end 
electronics.
Each sensor has 768 strips on each side with a pitch of 
95~$\mu$m. 
The stereo angle is 35~mrad, which is a compromise between stereo view and 
reduction of ambiguities resulting from high particle densities. 
The strips are almost parallel to
the beam axis ($z$-direction), to provide the best resolution in
the  $r\varphi$ direction. 

The modules are assembled on ladders of the same design as those supporting 
the SDD~\cite{giraudo}. A view of the SSD ladder is shown 
in Fig.~\ref{fig:ssdladder}.
The innermost SSD layer (layer 5) is composed of 34 ladders, each of them being
a linear array of 22 modules along the beam direction.
Layer 6 (the outermost ITS layer) consists of 38 ladders, each made of
25 modules.
In order to obtain full pseudo-rapidity coverage, the modules are mounted
on the ladders with small overlaps between successive modules, that are
staggered by
600~$\mu$m in the radial direction.
The 72 ladders, carrying a total of 1698 modules, are mounted on 
support cones in two cylinders. 
Carbon fiber is lightweight (to minimize the interactions) and at the
same time it is a stiff material allowing to minimize the
bending due to gravity, which is expected to 
give shifts of at most $50~\mum$, for the modules 
at the centre of the lateral ladders of the outer SSD layer.

For each layer, neighbouring ladders are mounted at one of two slightly 
different radii ($\Delta r=6~\mm$) such that full azimuthal 
coverage is obtained.
The acceptance overlaps, present both along $z$ and $r\varphi$, 
amount to 2\% of the SSD sensor surface.
The positions of the sensors with respect to reference points on the ladder
were measured during the detector construction phase, as well as the ones of 
the ladders with respect to the support cones.

The spatial resolution of the SSD system is determined by the 95~$\mu$m 
pitch of the sensor readout strips and the charge-sharing between 
those strips.
Without making use of the analogue information the r.m.s spatial resolution 
is 27~$\mu$m. Beam tests~\cite{SSDtestbeam} 
have shown that a spatial resolution of better than 
20~$\mu$m in the r$\varphi$ direction can be obtained by analyzing the 
charge distribution within each cluster. 
In the direction along the beam, the spatial resolution is of about 830~$\mu$m.

The SSD gain calibration has two components: overall calibration of ADC 
values to energy loss and relative calibration of the P and N sides. 
This charge matching is a strong point of double sided silicon sensors and 
helps to remove fake clusters. 
Both the overall and relative calibration are obtained from the data. 
Since the signal-to-noise ratio is larger than 20, the detection 
efficiency does not depend much on the details of the gain calibration.


\section{Alignment target and strategy}
\label{sec:strategy}

\noindent
For silicon tracking detectors, the typical target of the alignment 
procedures is to achieve a level of precision and accuracy such that 
the resolution on the reconstructed track parameters (in particular, 
the impact parameter and the curvature, which measures the transverse
momentum) is not degraded significantly with respect to the resolution expected
in case of the ideal geometry without misalignment. For the ALICE ITS,
this maximum acceptable degradation has been conventionally set to 20\%
(a similar target is adopted also for the ATLAS Inner 
Detector~\cite{atlasproc}). 
The resolutions on the track impact parameter and curvature are 
both proportional to the space point resolution, in the limit of negligible
multiple scattering effect (large momentum).
If the residual misalignment
is assumed to be equivalent to random gaussian spreads in the six alignment
parameters of the sensor modules, on which space points are measured,
a 20\% degradation in the effective space point resolution (hence 20\% degradation of the track parameters in the large momentum limit) 
is obtained when the misalignment
spread in a given direction is $\sqrt{120\%^2-100\%^2}\approx 70\%$ of the intrinsic sensor resolution along that direction.
With reference to the intrinsic precisions listed in Table~\ref{tab:ITSlayers},
 the target residual misalignment spreads in the local coordinates
 on the sensor plane are:
for SPD, $8~\mum$ in $x_{\rm loc}$ and $70~\mum$ in $z_{\rm loc}$; 
for SDD, $25~\mum$ in $x_{\rm loc}$ and $18~\mum$ in $z_{\rm loc}$; 
for SSD, $14~\mum$ in $x_{\rm loc}$ and $500~\mum$ in $z_{\rm loc}$.
Note that these spreads represent effective alignment spreads, including the significant effect of the $\theta_{\rm loc}$ angle (rotation about the 
axis normal to the sensor plane) on the spatial resolution.
In any case,
these target numbers are only an indication of the precision
that is required to reach an acceptable alignment quality. We will aim at 
getting even closer to the design performance expected in case
of ideal geometry.

The other
alignment parameters ($y_{\rm loc}$, $\psi_{\rm loc}$, $\varphi_{\rm loc}$) describe movements of the
modules mainly in the radial direction. These have a small impact on the 
effective resolution, 
for tracks with a small angle with respect to the normal to the module plane,
a typical case for tracks coming from the interaction region.
However, they are related to the so-called {\em weak modes}: correlated 
misalignments of the different modules that do not affect the 
reconstructed tracks fit quality ($\chi^2$), but bias
systematically the track parameters. A typical example is radial 
expansion or compression of all the layers, which biases the
measured track curvature, hence the momentum estimate. Correlated
misalignments for the parameters on the sensor plane 
($x_{\rm loc}$, $z_{\rm loc}$, $\theta_{\rm loc}$)
can determine
weak modes as well. 
These misalignments are, by definition, difficult to determine with 
tracks from collisions, but can be addressed using physical 
observables~\cite{lhcalignproc} 
(e.g. looking for shift in invariant masses of reconstructed 
decay particles) and cosmic-ray
tracks. These offer a unique possibility to correlate modules that are
never correlated when using tracks from the interaction region, and they 
offer a broad range of track-to-module-plane incidence angles that help to
constrain also the $y_{\rm loc}$, $\psi_{\rm loc}$ and $\varphi_{\rm loc}$ parameters, thus improving 
the sensitivity to weak modes.

As already mentioned in the introduction, the sources of alignment information
that we use are the survey measurements and the reconstructed space points from
cosmic-ray and collision particles. These points
are the input for the software alignment methods, based on global or
local minimization of the residuals. 

The general strategy for the ITS first alignment starts with the validation 
of the construction survey measurements of the SSD detector 
with cosmic-ray tracks and 
continues with the software alignment of the SPD and the SSD detectors, which
also uses cosmic-ray tracks, collected without magnetic field.
The initial alignment is more robust if 
performed with straight tracks (no field), which help to 
 avoid possible biases that can be introduced when working with curved tracks
(e.g. radial layer compression/expansion).
Then, the already aligned SPD and SSD are used to confirm and refine 
      the initial time-zero calibration of SDD, obtained with SDD standalone 
        methods. 
        These first steps are described in this report, which presents
        the status of the ITS alignment before the start of the LHC
        with proton--proton collisions.

The next step will be, after the       
validation of the SDD survey measurements with cosmic-ray tracks,
the alignment of the full detector (SPD, SDD, SSD) with tracks 
from cosmic rays and, mainly, from proton--proton collisions
collected with magnetic field $\rm B=0$ and $\rm B=0.5~T$. In particular,
    the data with magnetic field switched on will allow us 
to study the track quality and precision 
    as a function of the measured track momentum, thus separating
     the detector resolution and residual misalignment 
     from multiple scattering. The tracks from collisions
     will provide a uniform coverage of the detector modules and will also 
     be used
to routinely monitor the quality of the alignment during data taking,
and refine the corrections if needed.
The last step will be the relative alignment of the ITS and the TPC
with tracks, 
when both detectors will be internally aligned and calibrated. In addition,
      the relative movement of the ITS with respect to the TPC
      is being monitored, and to some extent measured, using a dedicated 
      system based on lasers, mirrors and cameras~\cite{itsams}.


\section{Cosmic-ray run 2008: data taking and reconstruction}
\label{sec:cosmics}


\noindent
During the 2008 cosmic run, extending from June to October, about $10^5$ events 
with reconstructed tracks in the ITS were collected.
In order to simplify the first alignment round, the solenoidal 
magnetic field was switched off during most of this data taking period.
The status of the three ITS sub-detectors during the data taking is summarized
in the following paragraph (for more details on the sub-detectors 
commissioning, see Refs.~\cite{spdcomm,sddcomm,ssdcomm}). 
The corresponding status during the 
first LHC runs with proton--proton collisions is given in Ref.~\cite{alice1st}.

For the SPD, 212 out of 240 modules (88\%) were active.
Noisy pixels, corresponding to less than 
0.15\% of the total number of pixels, were masked out, and the information 
was stored in the Offline 
Conditions Database (OCDB) to be used in the offline reconstruction.
For the SDD, 246 out of 260 modules (95\%) participated in the data 
acquisition. The baseline, gain and noise for each of the 133,000 anodes
were measured every 24 hours by means of dedicated calibration runs
that allowed us also to tag noisy ($\approx 0.5\%$) and dead (1\%) 
channels. The drift velocities were measured with dedicated injector runs collected every 6 hours, stored in the OCDB and
successively used in the reconstruction.
For the SSD, 1477 out of 1698 modules (87\%) were active.
The fraction of bad strips was $\approx 1.5\%$.
The normalized difference in P- and N-charge had a FWHM of 11\%.
The gains proved to be stable during the data taking.

The events to be used for the ITS alignment were collected with a
trigger provided by the pixel detectors (SPD).
The SPD FastOR trigger~\cite{aliceJINST} is based on a programmable hit 
pattern recognition system (on FPGA) 
at the level of individual readout chips (1200 in total, each reading a 
sensor area of about $1.4\times1.4~\cm^2$).
This trigger system allows for a flexible selection of 
events of interest, for example high-multiplicity proton--proton collisions, 
foreseen to be studied in the scope of the ALICE physics program.
For the 2008 cosmic run, the trigger logic consisted of selecting events
with at least one hit on the upper half of the outer SPD layer ($r\approx 7~\cm$)
and at least one on the lower half of the same layer. 
This trigger condition enhances significantly the 
probability of selecting events in which
a cosmic muon, coming from above (the dominant component of the cosmic-ray particles reaching the ALICE cavern
placed below $\approx30$~m of molasse),
traverses the full ITS detector.
This FastOR trigger is very efficient (more than 99\%) and has 
purity (fraction of events with a reconstructed track having points in 
both SPD layers)
reaching about 30--40\%, limited mainly by the radius of the inner layer 
($\approx 4~\cm$)
because the trigger assures 
only the passage of a particle through the outer layer ($\approx 7~\cm$). 
For the FastOR trigger, typically 77\% of the chips 
(i.e. about 90\% of the active modules) could be configured and used.  
The trigger rate was about 0.18~Hz.

The following procedure, fully integrated in the AliRoot framework~\cite{aliroot}, is used for track reconstruction.
After the  cluster finding in the ITS 
(hereafter, we will refer to the clusters as ``points''),
a pseudo
primary vertex is created using three aligned points 
in two consecutive layers (starting the search from the SPD). 
Track reconstruction 
is then performed using the ITS standalone tracker (as described in~\cite{crescio,PPR2}),
which finds tracks in the outward direction, from the innermost SPD layer to the outermost SSD layer,
using the previously found pseudo primary vertex as its seed; all found tracks
are
then refitted using the standard Kalman-filter fit procedure as implemented in the default ITS tracker.
During the track refit stage,
when the already identified ITS
points are used in the Kalman-filter fit in the inward direction, in order to obtain the track 
parameters estimate at the (pseudo) vertex,
 ``extra'' points 
are searched for in the ITS module overlaps.
For each layer, a search road for these overlap points in the 
neighbouring modules is defined
with a size of about seven times the current track position error.
Currently, the ``extra'' points
are not used to update the track parameters, so they can be exploited as a 
powerful tool to evaluate the ITS alignment quality.

\begin{figure}[!t]
\begin{center}
\includegraphics[width=0.49\textwidth]{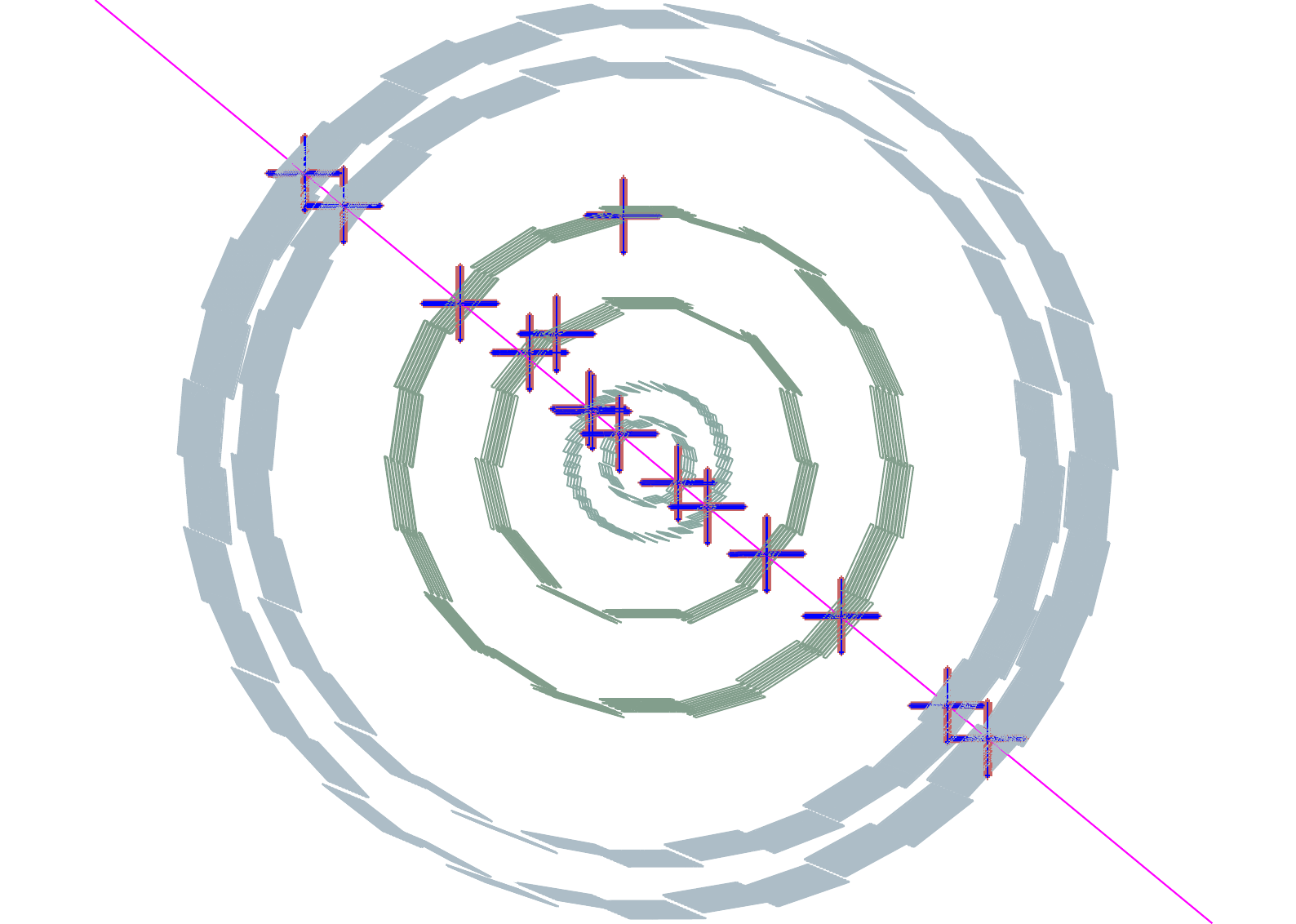}
\includegraphics[width=0.49\textwidth]{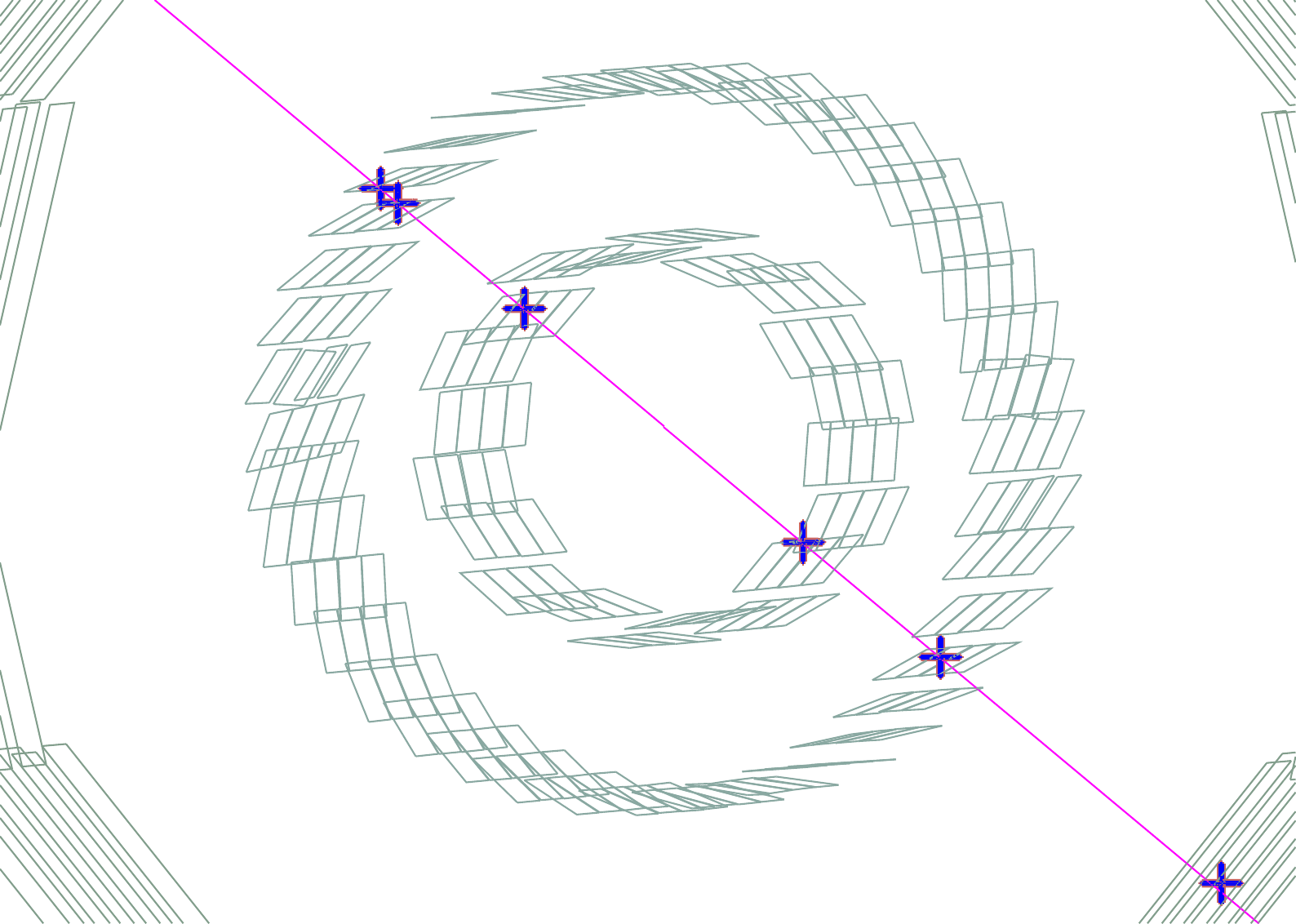}
\caption{(colour online) A clean cosmic event reconstructed in the ITS (left), as visualized
in the ALICE event display. The zoom on the SPD (right) shows an ``extra'' point
in one of the $r\varphi$ acceptance overlaps of the outer layer.}
\label{fig:display}
\end{center}
\end{figure}

A clean cosmic event consists of two separate tracks, 
one ``incoming'' in the top part of the ITS and one ``outgoing'' 
in the bottom part. Their matching at the
reference median plane ($y=0$) can be used as another alignment quality check. These two track halves are 
merged together in a single array of track points, which is the single-event 
input for the track-based alignment algorithms. 
A typical event of this type, as 
visualized in the ALICE event display, is shown in Fig.~\ref{fig:display}.

\begin{figure}[!t]
\begin{center}
\includegraphics[width=0.5\textwidth]{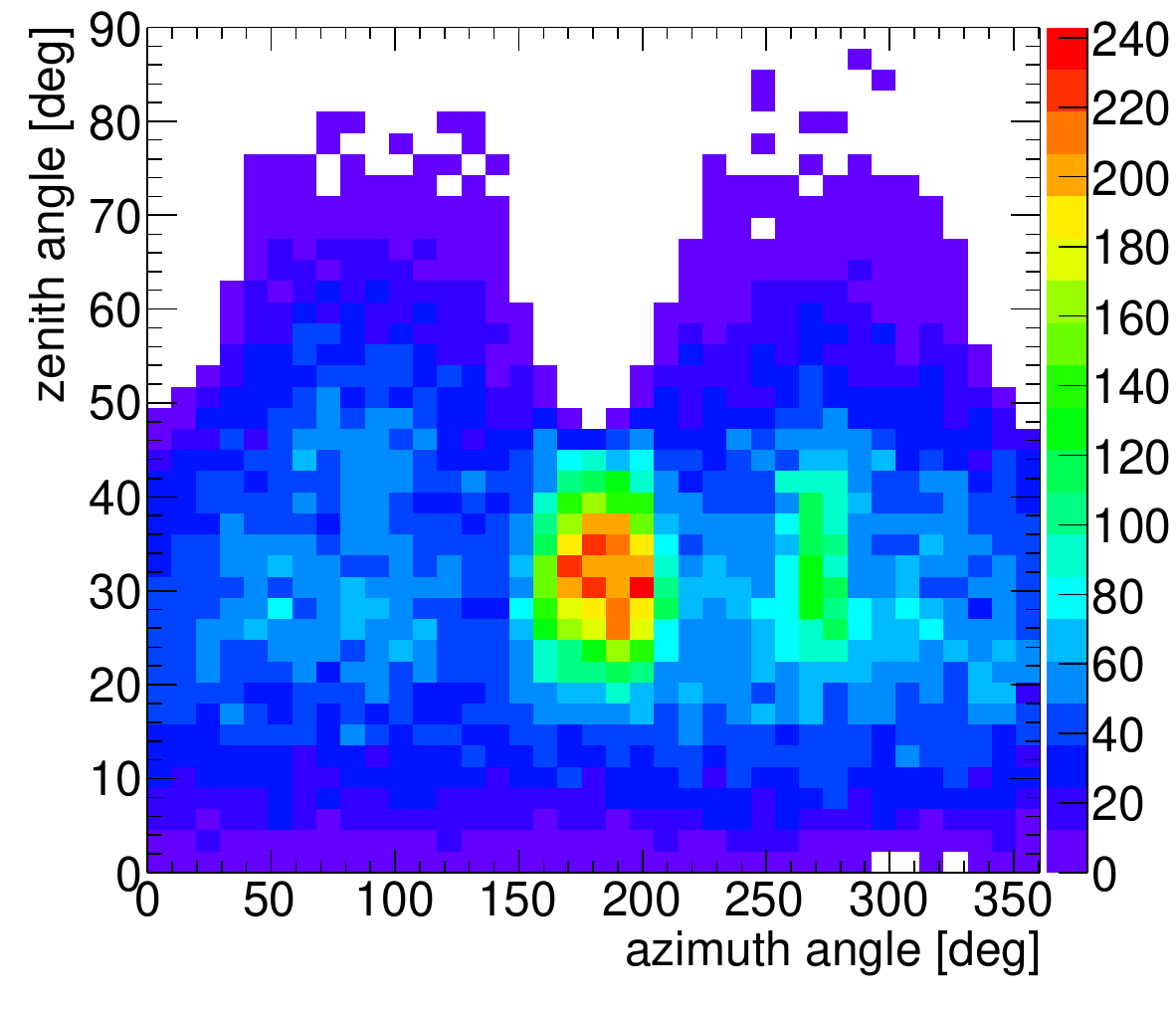}
\caption{(colour online) Uncorrected distribution of the zenith-azimuth angles
of the cosmic tracks reconstructed in the ITS.}
\label{fig:ZenithAzimuth}
\end{center}
\end{figure}

The uncorrected 
zenith-azimuth 2D distribution of the (merged) tracks with at least eight
points in the ITS  
is shown in Fig.~\ref{fig:ZenithAzimuth}, where the azimuth angle is defined
in a horizontal plane
starting from the positive side of the $z$ global axis.
The modulations in the azimuthal dependence of the observed flux are due to 
the presence of inhomogeneities in the molasse above the ALICE cavern, 
mainly the presence of two access shafts. These are seen as the structures
at zenith angle $\approx 30^\circ$ and azimuth $\approx 180^\circ$ (large shaft) 
and $\approx 270^\circ$ (small shaft).
On top of these structures, the effect of the SPD outer layer geometrical
acceptance is visible: the azimuthal directions perpendicular to the $z$
axis (around $90^\circ$ and $270^\circ$) 
have larger acceptance in the zenith angle.

The main limitation of the usage of cosmic-ray tracks
 for the alignment of a cylindrical 
 detector like the ITS is that
the occupancy
of the side modules (zenith angles approaching $90^\circ$) is small, especially for the external layers~\cite{noteITSalign}.
In the case of the SSD outer layer, which has the smallest 
fractional coverage, about 75\% of the ladders are covered.
This is due to   
 the small size of the triggering detector (SPD), 
the dominance of small zenith angles for cosmic-ray particles 
and the cut on the track-to-module incidence angle ($>30^\circ$) that
we apply to reject large and elongated clusters.


\section{Validation of the survey measurements with cosmic-ray tracks}
\label{sec:survey}


\noindent
The SDD and SSD were surveyed during the assembling phase
 using a measuring machine.
The survey, very similar for the two detectors,
 was carried out in two stages: the measurement of the
positions of the modules 
on the ladders and the measurement of the positions of the
ladder end points on the support cone. 

In the first stage, for SDD for example~\cite{noteSDDsurvey},
the three-dimensional positions of six reference markers
engraved on the detector surface were measured for each module
with respect to ruby reference spheres fixed to the support structure.
The precision of the measuring machine
was 5~$\mum$ in the coordinates on the ladder plane and
about 10~$\mum$ in the direction orthogonal to the plane.
The deviations of the reference marker coordinates on the plane
with respect to design positions showed
an average value of 1~$\mum$ and a r.m.s. of 20~$\mum$.
In the second stage, the positions of the ladder end points with respect to
the cone support structure were measured with a precision of
about 10~$\mum$. However, for the outer SSD layer,
the supports were dismounted and remounted
after the survey; the precision of the remounting procedure is
estimated to be around 20~$\mum$ in the $r\varphi$ direction~\cite{aliceJINST}.

In the following we describe the results for the 
validation of the SSD survey measurements with cosmic-ray data.
The validation of the SDD survey will be performed after completion of 
the detector calibration.

\subsection{Double points in SSD module overlaps}

\noindent
As already mentioned, 
the modules are mounted with a small (2~mm) overlap for
both the longitudinal ($z$, modules on the same ladder) and transverse
directions ($r\varphi$, adjacent ladders). These overlaps allow us to
verify the relative position of neighbouring modules using double points
produced by the same particle on the two modules. 
Since the two points
are very close in space and the amount of material crossed by the particle 
between the
two points is very limited, multiple scattering can be neglected.

\begin{figure}[!t]
\begin{center}
\includegraphics[width=0.49\textwidth]{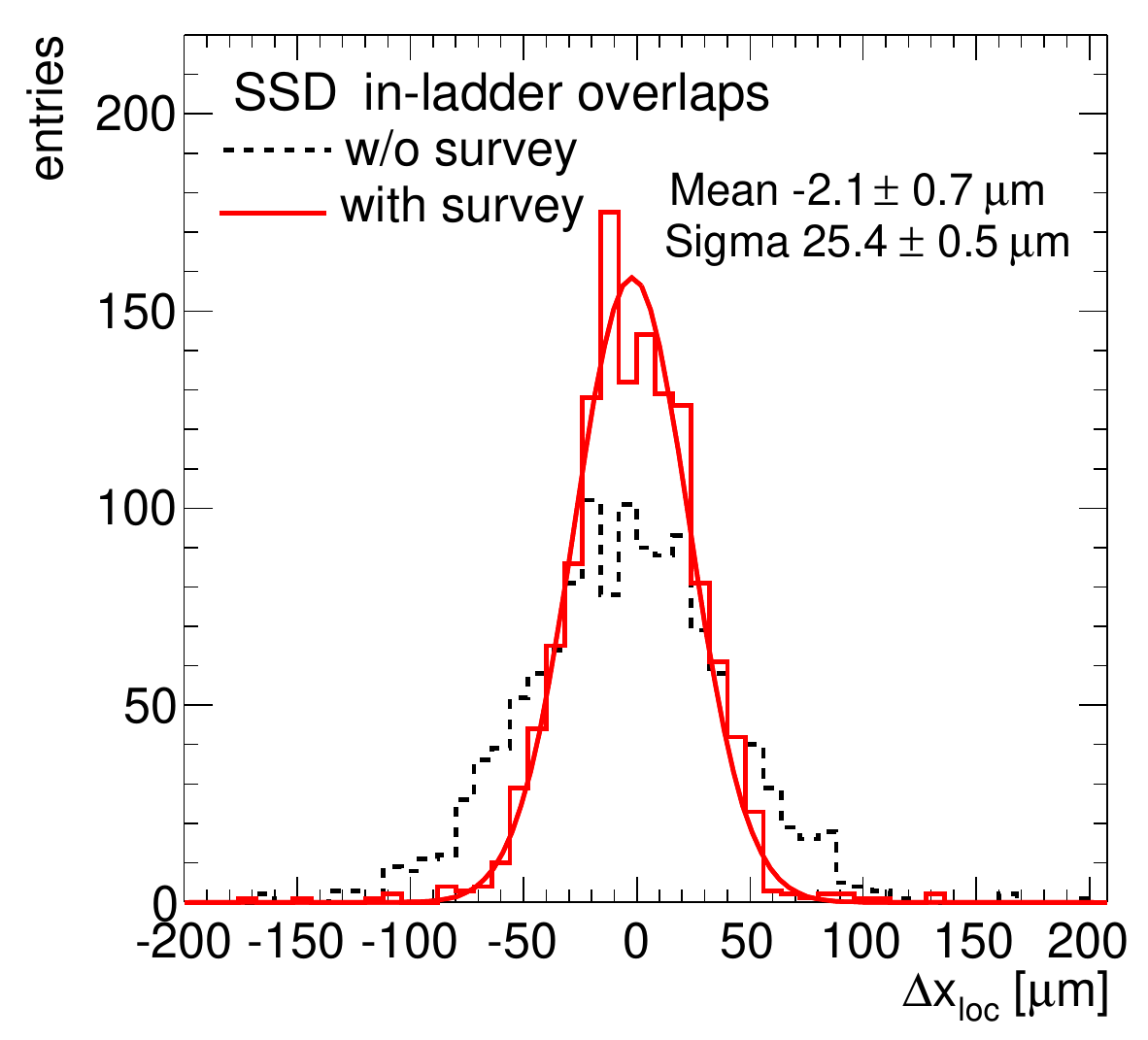}
\includegraphics[width=0.49\textwidth]{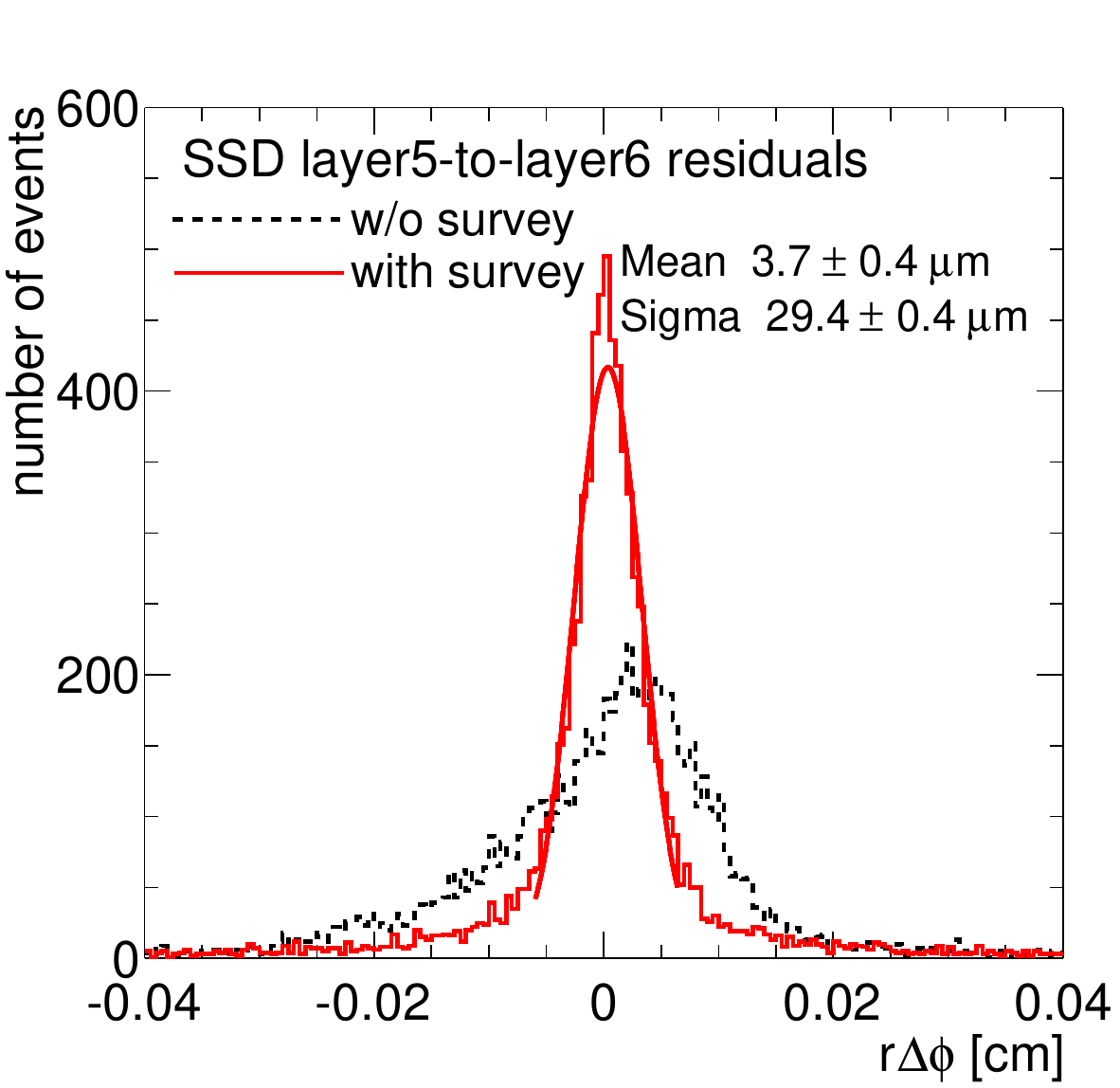} 
\caption{(colour online) SSD survey validation. Left: distribution of $\dxloc$, the
  distance between two points in the module overlap regions along $z$ on the
  same ladder. 
  Right: distribution of the $r\varphi$ 
  residuals between straight-line tracks defined from two points on 
  layer 6 and the corresponding points on layer 5. 
  In both cases, gaussian fits to the distributions with survey applied are shown
  (in right-hand panel the fit range is $\pm2\,\sigma$, i.e. 
  $[-60~\mum,+60~\mum]$).}
\label{fig:ssdsurvey}
\end{center}
\end{figure}

We define the distance $\dxloc$ 
between the two points in the local $x$ direction on 
the module plane ($\approx r\varphi$) by projecting, along the track 
direction, the point of one of the two modules on the other module plane.
Figure~\ref{fig:ssdsurvey} (left) shows the $\dxloc$ distribution
with and without the survey corrections,
for both SSD layers (it was verified that the distributions for the 
two layers are compatible~\cite{noteITSalign,noteSSDsurvey}) . 
When the survey corrections are applied, 
the spread of the distributions, obtained from 
a gaussian fit, is $\sigma\approx 25.5~\mum$. This arises from
the combined spread of the two points, thus the corresponding
effective position resolution
for a single point is estimated to be smaller by a factor $1/\sqrt{2}$, 
i.e. $\approx 18~\mum$, which is compatible with the expected 
intrinsic spatial resolution of about $20~\mum$.
This indicates that the residual misalignment after
applying the survey is negligible with respect to the
intrinsic spatial resolution. 
This validation procedure was confirmed using Monte Carlo 
simulations of cosmic muons in the
detector without misalignment, which give a spread in
$\dxloc$ of about $25~\mum$, in agreement with that obtained from the data.

\subsection{Track-to-point residuals in SSD}

\noindent
Another test that was performed uses two points in the outer SSD layer
to define a straight track (no magnetic field) and inspects the
residuals between points on the inner layer and the track.
The residuals are calculated using the position along the track 
corresponding to the minimum of the weighted (dimensionless) distance to the 
point\footnote{The different
expected resolutions in $r\varphi$ and $z$ have been taken into account
in the calculation of the distance of closest approach by dividing
the deviations by the expected uncertainties, i.e. making use of
a dimensionless distance measure.}.
Figure~\ref{fig:ssdsurvey} (right) shows the distribution of the
$r\varphi$ residuals 
between tracks through layer 6 and points on layer 5. 
The distribution exhibits significant non-gaussian tails, 
due to multiple scattering of low-momentum particles. The effect of multiple
scattering on the residuals was analytically estimated to be of about
$300~\mum$ for $p=0.5~\gev/c$ and to be negligible for $p\gsim 7~\gev/c$.
This is roughly compatible with the observed distribution of residuals.
The width of the central part of the distribution is quantified by
performing a gaussian fit truncated at $2\,\sigma$, 
that gives $\sigma_{r\varphi} = 29~\mum$.
The spread contains a contribution from
the uncertainty in the track trajectory due to the uncertainties in
the points on the outer layer. Assuming the same resolution on the
outer and inner layers and taking into account the geometry of the
detector, the effective single point resolution spread is
$1/\sqrt{1.902}$ times the overall spread~\cite{noteSSDsurvey}, 
that is $21~\mum$.
This spread is larger than the effective resolution of about $18~\mum$
that is extracted from the double points in module overlaps.
This difference could be partly due to the multiple scattering, 
relevant for this analysis and negligible for the overlaps analysis, but
we can not rule out
that additional misalignments with a r.m.s. up to about $15~\mum$ are
present in the SSD. The mean residual is also non-zero, $(3.7 \pm 0.4)~\mum$, which suggests that residual shifts at the 5--$10~\mum$ level
could be present. These misalignments would have to be at the ladder
level to be compatible with the result from the study with sensor
module overlaps.

The same analysis was performed for the residuals in the $z$ 
direction~\cite{noteITSalign,noteSSDsurvey}, 
not shown here. The distributions without
and with survey were found to be compatible and the corresponding 
effective single point resolution was found to be compatible
with the expected intrinsic resolution of about $800~\mum$.
This indicates that the residual misalignment in $z$ is much smaller than the
intrinsic SSD resolution.

A third method that was used to verify the SSD survey consisted in
performing tracking with pairs of points (2 points on layer 5 and two
points on layer 6 or two sets of points on layer 5 and 6), and
comparing the track parameters of both track segments. The conclusion
from this method is consistent with the results from the
track-to-point method. For details see~\cite{noteSSDsurvey}.


\section{ITS alignment with Millepede}
\label{sec:millepede}

\noindent 
In general, the task of the track-based alignment algorithms 
is determining
the set of geometry parameters that minimize the global 
$\chi^2$ of the track-to-point residuals:
\begin{equation}
\chi^2_{\rm global} = \sum_{\rm modules,\,tracks}{ \vec{\delta}_{t,p}^{\rm\,T} \, {\bf V}^{-1}_{t,p} \, \vec{\delta}_{t,p} }\,.
\label{eq:chi2}
\end{equation}
In this expression, the sum runs over all the detector modules and all 
the tracks in a given data set; 
$\vec{\delta}_{t,p} = \vec{r}_{t} - \vec{r}_{p}$
is the residual between the data point $\vec{r}_{p}$
and the reconstructed track extrapolation $\vec{r}_{t}$ to
the module plane; ${\bf V}_{t,p}$ is the covariance matrix
 of the residual. Note that, in general, the reconstructed tracks
themselves depend on the assumed geometry parameters.
This section describes how this minimization problem is treated by
Millepede~\cite{refMP1,refMP2} ---the main algorithm used for ITS
alignment--- and presents the first alignment results
obtained with cosmic-ray data. 

\subsection{General principles of the Millepede algorithm}
\label{sec:mp_principle}

\noindent
Millepede belongs to the {\em global least-squares minimization} type of
algorithms, which aim at determining simultaneously all the parameters
that minimize the global $\chi^2$ in Eq.~(\ref{eq:chi2}).
It assumes that, for each of the local coordinates, the residual
of a given track $t$ to a specific
measured point $p$ can be represented in a linearized form as
$\delta_{t,p} = \vec{a} \cdot \partial \delta_{t,p}/\partial \vec{a} +
\vec{\alpha_{t}} \cdot \partial \delta_{t,p}/\partial \vec{\alpha_{t}}$, where
$\vec{a}$ is the set of global parameters describing the
alignment of the detector (three translations and three rotations per module) 
and $\vec{\alpha}_t$ is the set of local
parameters of the track.
The corresponding $\chi^{2}_{\rm global}$ equation for $n$
tracks with $\nu$ local parameters per track and for $m$ modules 
with 6 global parameters ($N=6\,m$ total global parameters)
leads to a huge set of $N+\nu\,n$ normal equations. 
The idea behind the Millepede method is to consider the local $\vec{\alpha}$
parameters as nuisance parameters that are eliminated using the 
Banachiewicz identity~\cite{invPart} for partitioned matrices.
This allows to build explicitly only
the set of $N$ normal equations for the global parameters.
 If needed, linear constraints on the
global parameters can be added using the Lagrange
multipliers. Historically, two versions, Millepede and Millepede II,
were released. The first one was performing
the calculation of the
residuals, the derivatives and the final matrix elements as well as 
the extraction of the
exact solution in one single step, 
keeping all necessary information in computer memory. The large
memory and CPU time needed to extract the exact solution of a $N\times
N$ matrix equation effectively limited its use to $N<10,000$ global
(alignment) parameters.
This limitation was removed in the second version, Millepede~II, which
builds the matrices (optionally) in sparse format, to save
memory space, and solves them using
advanced iterative methods, much faster than the
exact methods.

\subsection{Millepede for the ALICE ITS}

Following the development of Millepede, ALICE had its own
implementation of both versions, hereafter indicated as MP and MPII, within the
AliRoot framework~\cite{aliroot}. Both
consist of a detector independent solver class, responsible for building and
solving the matrix equations, and a class interfacing the
former to specific detectors. While MP closely
follows the original algorithm~\cite{refMP1}, MPII has a number  
of extensions. In addition to the MinRes matrix equation solution algorithm offered by the original
Millepede~II, the more general FGMRES~\cite{refFGMRES} method was added,
as well as the powerful ILU(k) matrix preconditioners~\cite{refILUk}. 
All the results shown in this work are obtained with MPII.

The track-to-point
residuals, used to construct the global $\chi^2$,
 are calculated using a parametric straight line $\vec{r}(t) =
\vec{a} + \vec{b}\,t$ or helix $\vec{r}(t) = \{a_x + r\,\cos(t+\varphi_0),
 a_y + r\,\sin(t+\varphi_0), a_z + b_z\,t \}$ track model, depending on the
presence of the magnetic field. 
The full error matrix of the measured points is accounted for in the 
track fit, 
while multiple scattering is ignored,
since it has no systematic effect on the residuals.

Special attention was paid to the possibility to account for
the complex hierarchy of the alignable volumes of the ITS,
in general leading to better description of the material budget distribution after alignment.
This is achieved by defining explicit parent--daughter 
relationships between the volumes corresponding to mechanical degrees of freedom in the ITS.
The alignment is performed simultaneously for the
volumes on all levels of the hierarchy, e.g. for the SPD the
corrections are obtained in a single step for the sectors, the
half-staves within the sectors and the modules
within the
half-staves. Obviously, this leads to a degeneracy of the possible
solutions, which should be removed by an appropriate set of
constraints. 
We implemented the possibility to constrain either the mean or the median 
of the corrections for the 
daughter volumes of any parent volume. While the former can be applied via
Lagrange multipliers directly at the minimization stage, the latter,
being non-analytical, is applied after the Millepede minimization in a special post-processing step.
The relative movement $\delta$ of volumes for which the survey data is available (e.g. SDD and SSD modules) can be
restricted to be within the declared survey precision $\sigma_{\rm survey}$ by adding a set of gaussian constraints $\delta^{2}/\sigma^{2}_{\rm survey}$ to the global $\chi^{2}$.
%

\begin{figure}[!t]
\centering
\includegraphics[width=0.49\textwidth]{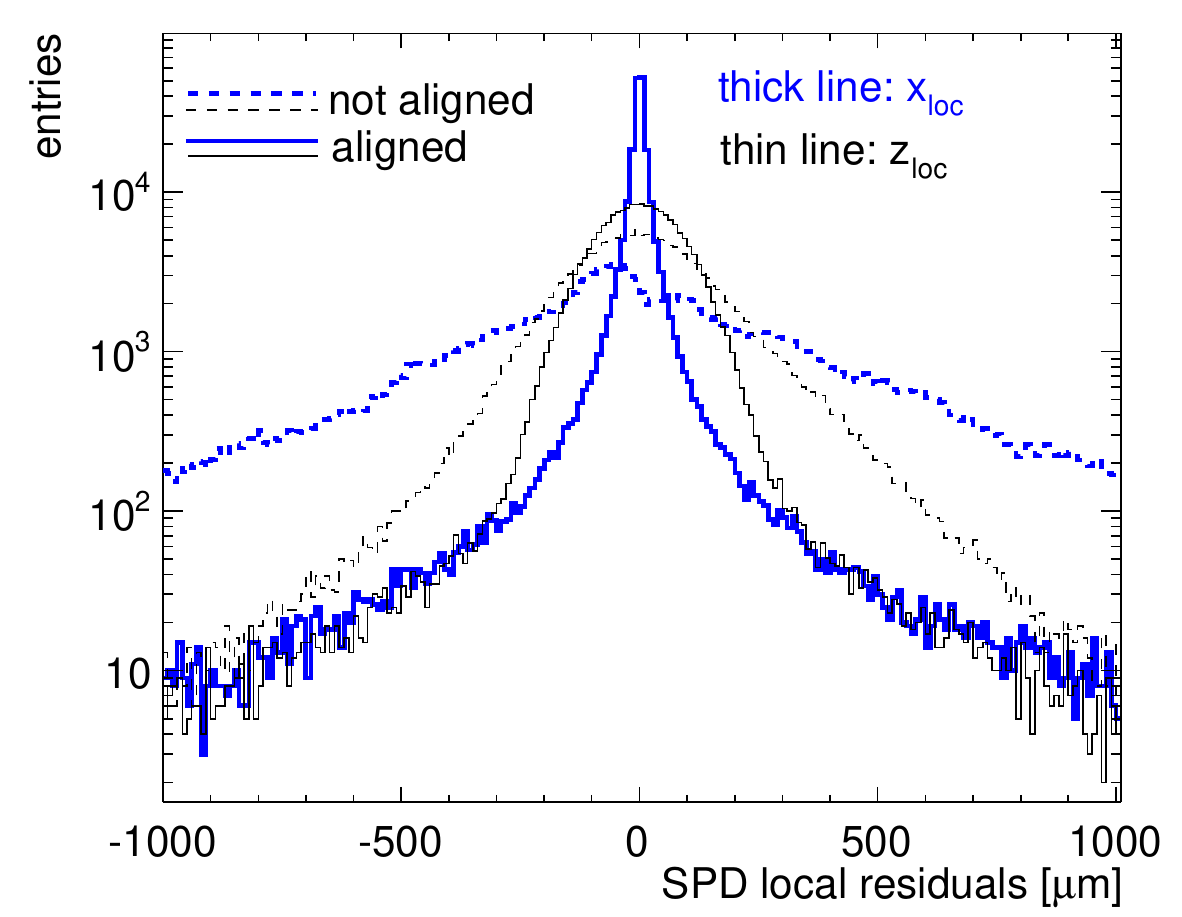}
\includegraphics[width=0.49\textwidth]{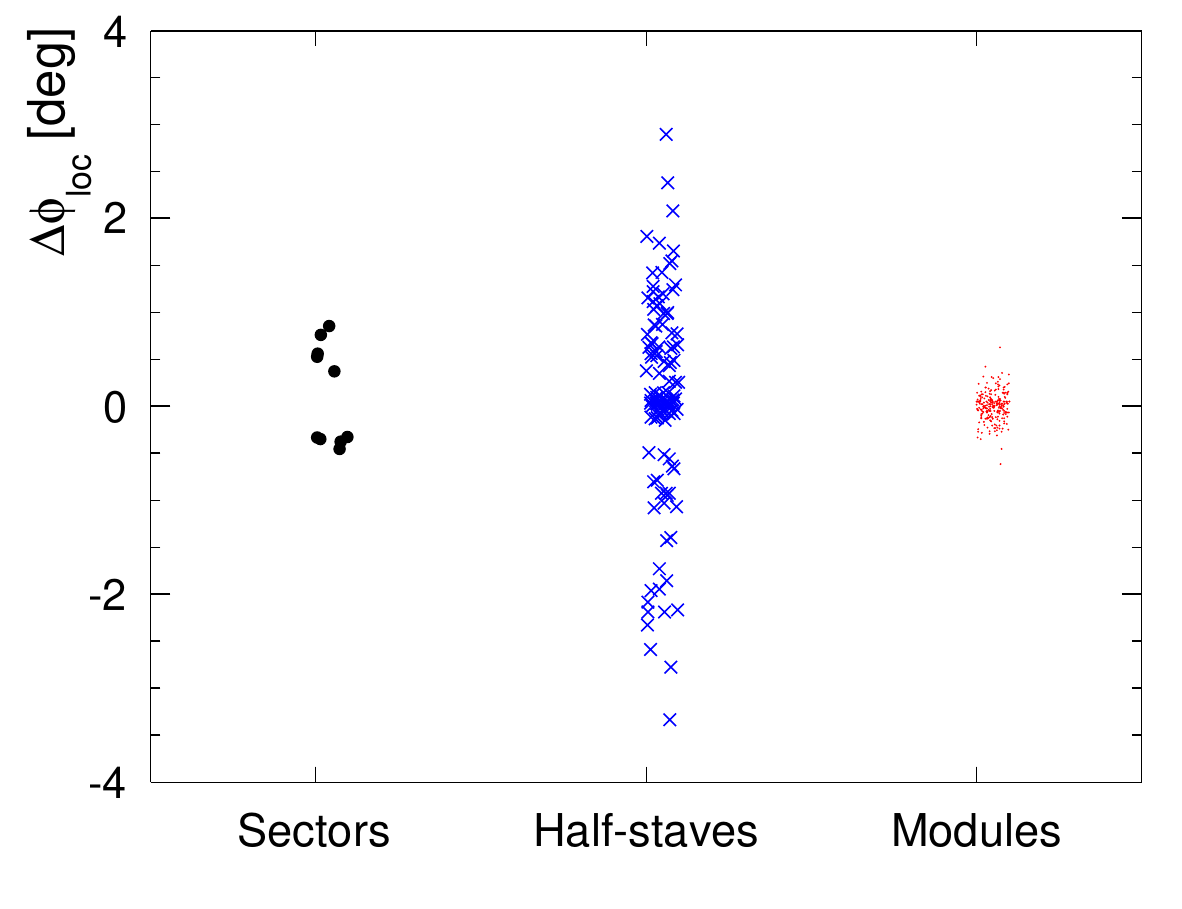}
\caption{(colour online) Left: example of Millepede residuals in the local reference frame of the SPD modules before and after the alignment.
Right: the corrections to the $\varphi_{\rm loc}$ angle obtained in the hierarchical SPD alignment with Millepede.}
\label{fig:aligSPD}
\end{figure}

We report here two example figures to illustrate the bare output results 
from Millepede~II for the alignment of the SPD detector (in this case),
while the analysis of the alignment quality will be 
presented in the next section. 
The left-hand panel of Fig.~\ref{fig:aligSPD} 
shows an example of the residuals in
SPD (in the local reference frame of the modules) before and after alignment.
The right-hand panel of Fig.~\ref{fig:aligSPD} shows 
the obtained corrections for the $\varphi_{\rm loc}$ angle (rotation of the
volume with respect to its $z_{\rm loc}$ axis), indicating that the 
largest misalignments are at the level of the half-staves with respect to 
the carbon fiber support sectors.


\subsection{Results on alignment quality}
\label{sec:milleresults}

\noindent
The SPD detector was first aligned using $5 \times 10^4$ cosmic-ray tracks, 
with two points in the inner layer and two points in the outer layer,
 collected in 2008 with the magnetic field switched off. 
As described in the previous section, the hierarchical alignment procedure 
consisted in:
aligning the ten sectors with respect to each other, the 
twelve half-staves of each sector with respect to the sector, and
the two modules of each half-stave with respect to the half-stave.
 
The following two observables are mainly used to check the quality 
of the obtained alignment: the top half-track to bottom half-track
matching at the plane $y=0$, and the track-to-point distance 
for the ``extra'' points in the acceptance overlaps.

For the first observable, the cosmic-ray track is split into the 
two track segments that cross the upper ($y>0$) and lower ($y<0$) 
halves of the ITS barrel, and the parameters of the two segments 
are compared at $y=0$. The main variable is $\dxy$, 
the track-to-track distance at $y=0$ in the $(x,y)$ 
plane transverse to beam line. 
This observable, that is accessible only with cosmic-ray tracks, 
provides a direct measurement of the 
resolution on the track transverse impact parameter $d_0$; 
namely: $\sigma_{\dxy}(\pt)=\sqrt{2}\,\sigma_{d_{0}}(\pt)$.
Since the data used for the current analysis were collected without
magnetic field, they do not allow us to directly assess the $d_0$
resolution (this will be the subject of a future work). 
However, also without a momentum measurement, $\dxy$ is a powerful indicator
of the alignment quality, as we show in the following. 

\begin{figure}[!t]
\centering
\includegraphics[width=0.49\textwidth]{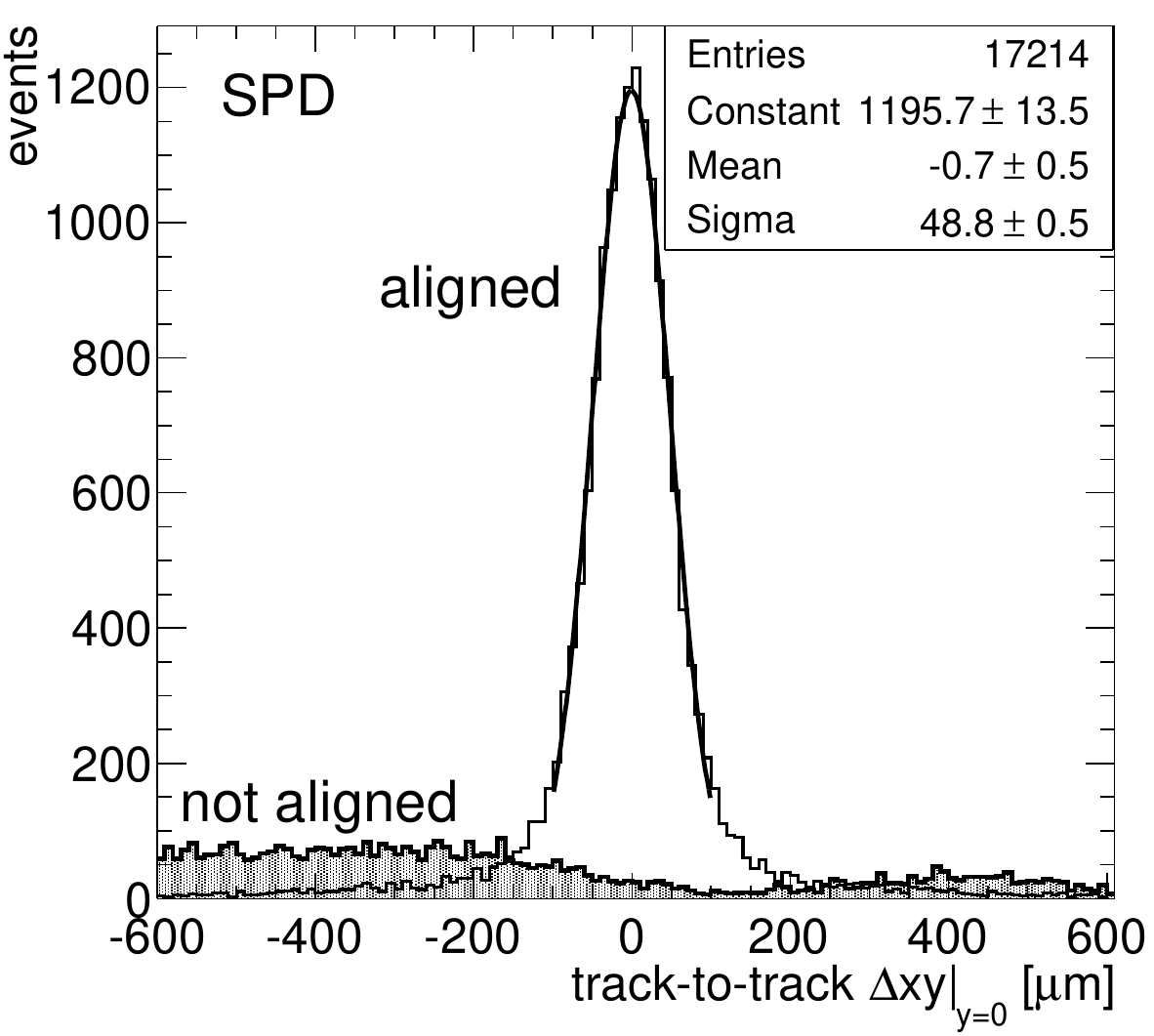}
\includegraphics[width=0.49\textwidth]{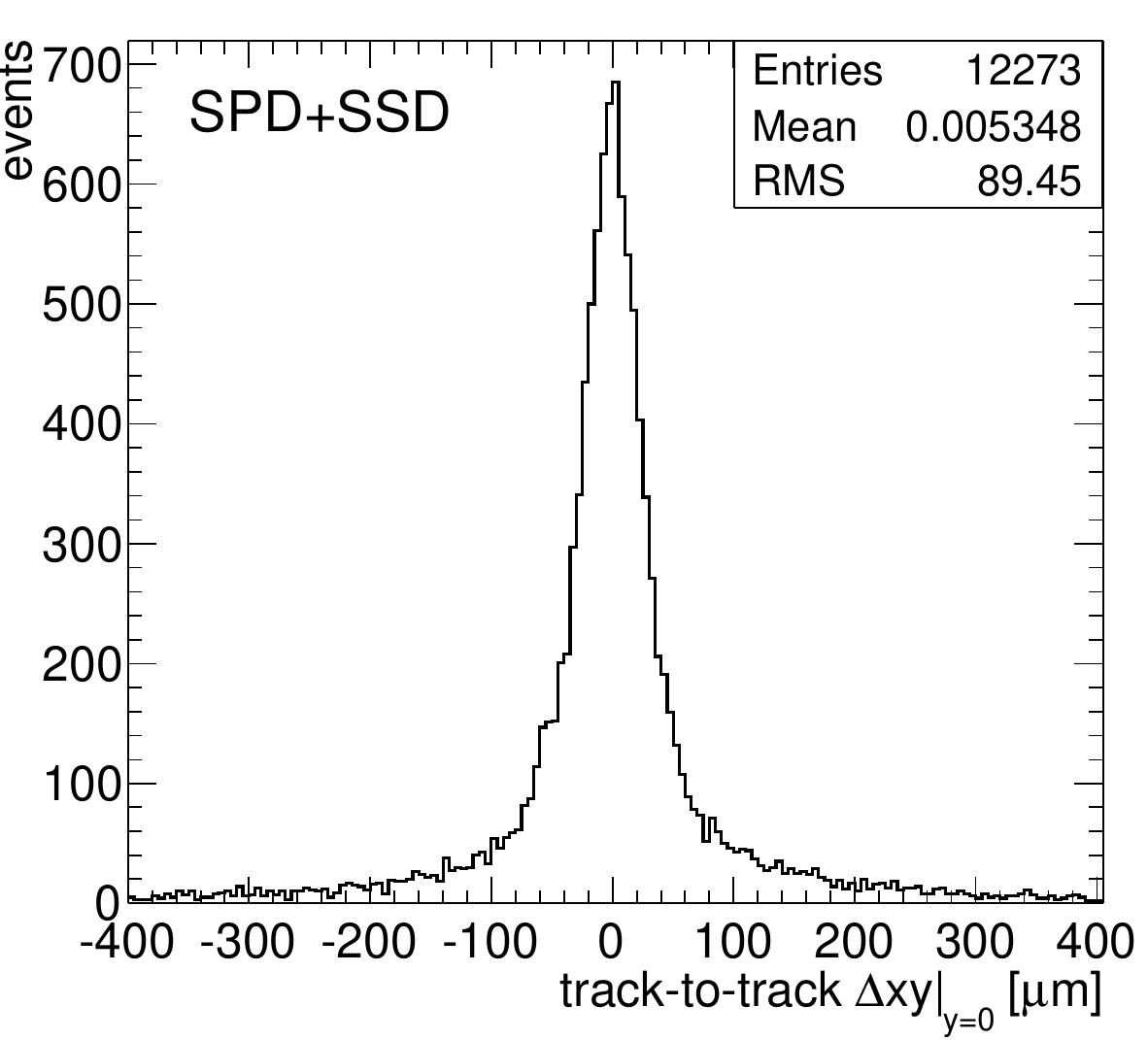}
\caption{Left: distribution of $\dxy$ for SPD only, before and after alignment.
Right: distribution $\dxy$ 
for track segments reconstructed in the upper and lower parts of SPD+SSD layers;
each track segment is required to have four assigned points;
SSD survey and Millepede alignment corrections are applied. In both cases,
the distributions are produced from the sample of events used to 
obtain the alignment corrections.}
\label{fig:deltaxy0}
\end{figure}

Figure~\ref{fig:deltaxy0} (left) shows the distribution of $\dxy$ 
for SPD, without and with the alignment corrections.
The two track segments are required to have a point in each of the SPD layers
and to pass, in the transverse plane, within 1~cm from the origin (this
cut selects tracks with a similar topology to those produced in collisions
and rejects tracks that have small incidence angles on the
inner layer modules). 
 A gaussian fit to the distribution 
in the range $[-100~\mum,+100~\mum]$ gives a centroid compatible with zero and a
spread $\sigma\approx50~\mum$. For comparison, a spread of $38~\mum$ is 
obtained from a Monte Carlo simulation, with the ideal geometry of the ITS
(without misalignment), 
of cosmic muons generated according to the momentum spectrum measured by the 
ALICE TPC in cosmic runs with magnetic field.
When only the SPD detector is used and the tracks are straight lines 
(no magnetic field), the spread of the $\dxy$ distribution 
can be related in a simple way to 
the effective spatial resolution $\sigma_{\rm spatial}$, which includes
the intrinsic sensor resolution and of the residual misalignment.  
For tracks passing close to the beam line (as in our case, with the cut
at $1~\cm$), we have:
\begin{equation}
\sigma_{\dxy}^{2}\approx
2\,\frac{(r^2_{\rm SPD1}\sigma^2_{\rm spatial,SPD1}+r^2_{\rm SPD2}\sigma^2_{\rm spatial,SPD2})}{(r_{\rm SPD1}-r_{\rm SPD2})^2}\approx 
2\,\frac{r^2_{\rm SPD1}+r^2_{\rm SPD2}}{(r_{\rm SPD1}-r_{\rm SPD2})^2}\,\sigma^2_{\rm spatial}\,, 
\end{equation}
where the inner and outer SPD layers are indicated as SPD1 and SPD2, 
respectively. This relation neglects the effect of multiple scattering 
in the pixels and in the beam pipe, which is certainly one of the reasons
why the $\dxy$ distribution is not gaussian outside the central region,
most likely populated by the high-momentum component of the cosmic muons.
Using the fit result, $\sigma_{\dxy}\approx 50~\mum$, 
obtained in the central region $[-100~\mum,+100~\mum]$, we estimate
the value $\sigma_{\rm spatial}\approx 14~\mum$, not far from the intrinsic
resolution of about $11~\mum$ extracted from the simulation.
However, 
a precise estimation of the effective spatial resolution with this method 
requires the measurement of the track momentum, 
to account properly for the multiple scattering contribution. 
The statistics collected in 2008 with magnetic field did not allow a 
momentum-differential analysis.

The next step in the alignment procedure is the inclusion of the SSD
detector. As shown in section~\ref{sec:survey}, the survey measurements
already provide a very precise alignment, with residual misalignment 
levels of less than $5~\mum$ for modules on the ladder
and of about $20~\mum$ for ladders.
Because of the limited available statistics ($\approx2\times10^4$ 
tracks with four points in SPD and four points in SSD), the
expected level of alignment obtained with Millepede on single SSD modules 
is
significantly worse than the level reached with the survey measurements. 
For this reason,
Millepede was used only to align the whole SPD barrel 
with respect to the SSD barrel and to optimize the positioning of
large sets of SSD modules, 
namely the upper and lower halves of layers 5 and 6. For this last step, 
the improvement on the global positioning of the 
SSD layers was verified by comparing the position and direction of the 
pairs of SSD-only track segments built using: two points in the upper
and two in the lower half-barrel (upper--lower configuration) or
two points in the inner and two in the outer layer (inner--outer configuration).
Before the alignment (only the survey corrections applied), the mean 
of $\dxy$ is $(120\pm7)~\mum$ and $(-1.8\pm0.6)~\mum$ for the upper--lower
and inner--outer configurations, respectively. After the alignment, 
it is $(-5\pm6)~\mum$ and $(0.5\pm0.6)~\mum$, respectively, that is, 
compatible with zero for both configurations.

The right-hand panel of Fig.~\ref{fig:deltaxy0} shows
the distribution of $\dxy$ 
for pairs of track segments, each 
reconstructed with two points in SPD and two in SSD, i.e. the merged
cosmic-ray track has eight points in SPD+SSD.
It can be seen that, when the SSD survey and the Millepede alignment 
are applied,
the distribution is centred at zero and very narrow 
($\rm FWHM \approx 60~\mum$), but it shows non-gaussian tails, 
most likely due to multiple scattering.
A more precise alignment of the
SSD using high-momentum
tracks will be performed with the 2009 cosmic-ray and proton--proton 
data.

The second alignment quality observable is the $\dxloc$ distance between 
points in the region where there is an acceptance overlap between two
modules of the same layer.
Because of the short radial distance between the two overlapping 
modules (a few mm), 
the effect of multiple scattering is negligible. However, in order to relate
the spread of $\dxloc$ to the effective resolution, the dependence of the 
intrinsic sensor resolution on the track-to-module incidence angle 
has to be accounted for. In particular, 
for SPD, due to the geometrical layout of the detector 
(Fig.~\ref{fig:spdsector}, left),
the track-to-module incidence angles in the transverse plane 
are in general not equal to $90^\circ$
and they are very different for two adjacent overlapping modules
crossed by the same track.
If $\dxloc$ is defined as described in section~\ref{sec:survey},
the error on $\dxloc$ can be related to the effective spatial resolution 
of the two points, $\sigma_{\rm spatial}$, as:
\begin{equation}
\sigma_{\dxloc}^2= \sigma_{\rm spatial}^{2}(\alpha_2)+ \sigma_{\rm spatial}^{2}(\alpha_1)\cos^2(\varphi_{12})
\label{sigma_ovl}
\end{equation}
where the 1 and 2 subscripts indicate the two overlapping points,
 $\alpha_i$ is the incidence angle of the track on the module plane, 
and $\varphi_{12}$ is the relative angle between the two module planes,
which is $18^\circ$ and $9^\circ$ on the inner and outer SPD layer, respectively.
Note that, for SSD overlaps on the same ladder, 
we have $\alpha_1=\alpha_2\simeq 90^\circ$ and
$\varphi_{12}=0$; therefore, $\sigma_{\dxloc}=\sqrt{2}\,\sigma_{\rm spatial}$, 
which is the relation we used in section~\ref{sec:survey}.

\begin{figure}[!t]
\begin{center}
\includegraphics[width=0.49\textwidth]{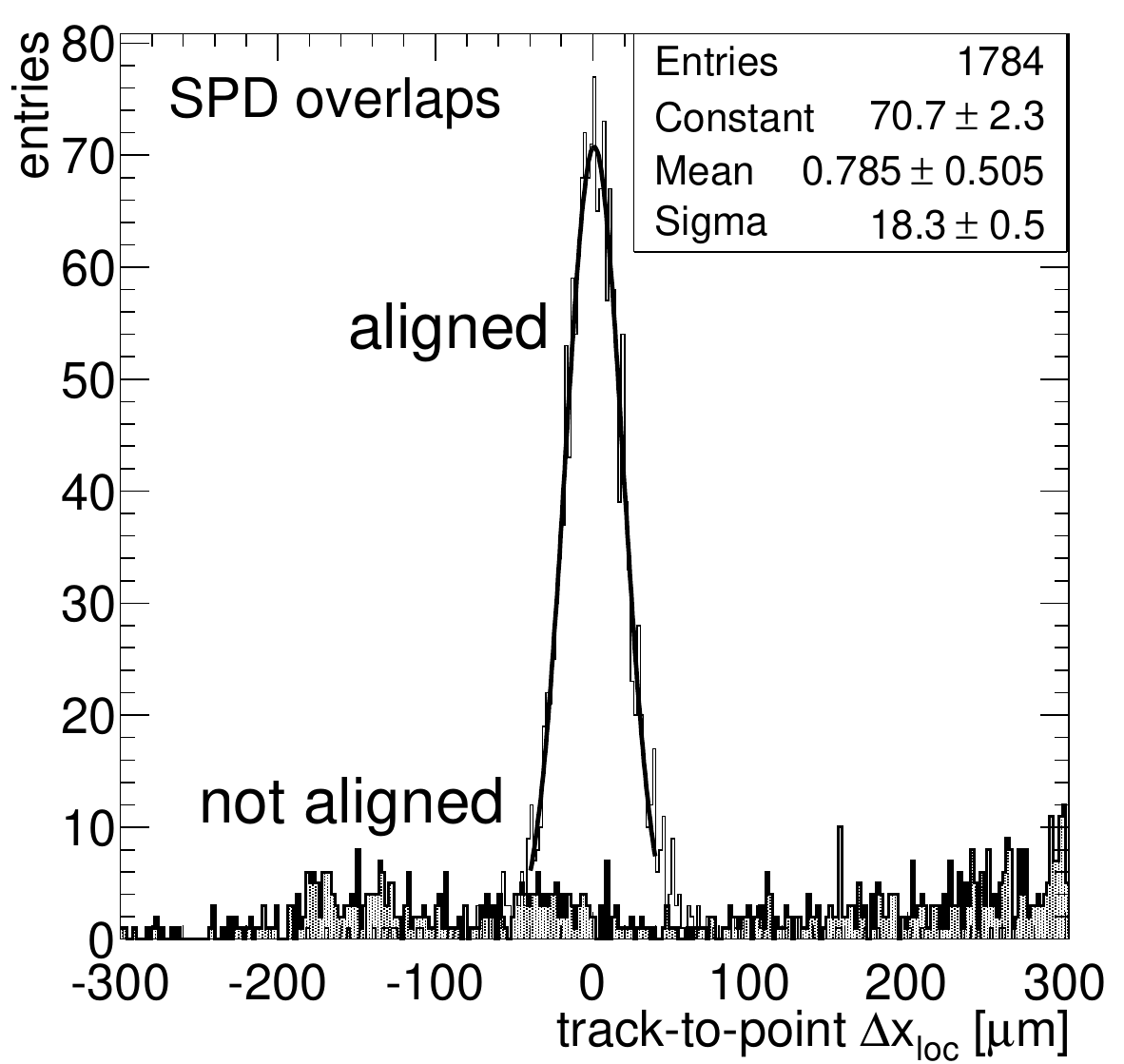}
\includegraphics[width=0.49\textwidth]{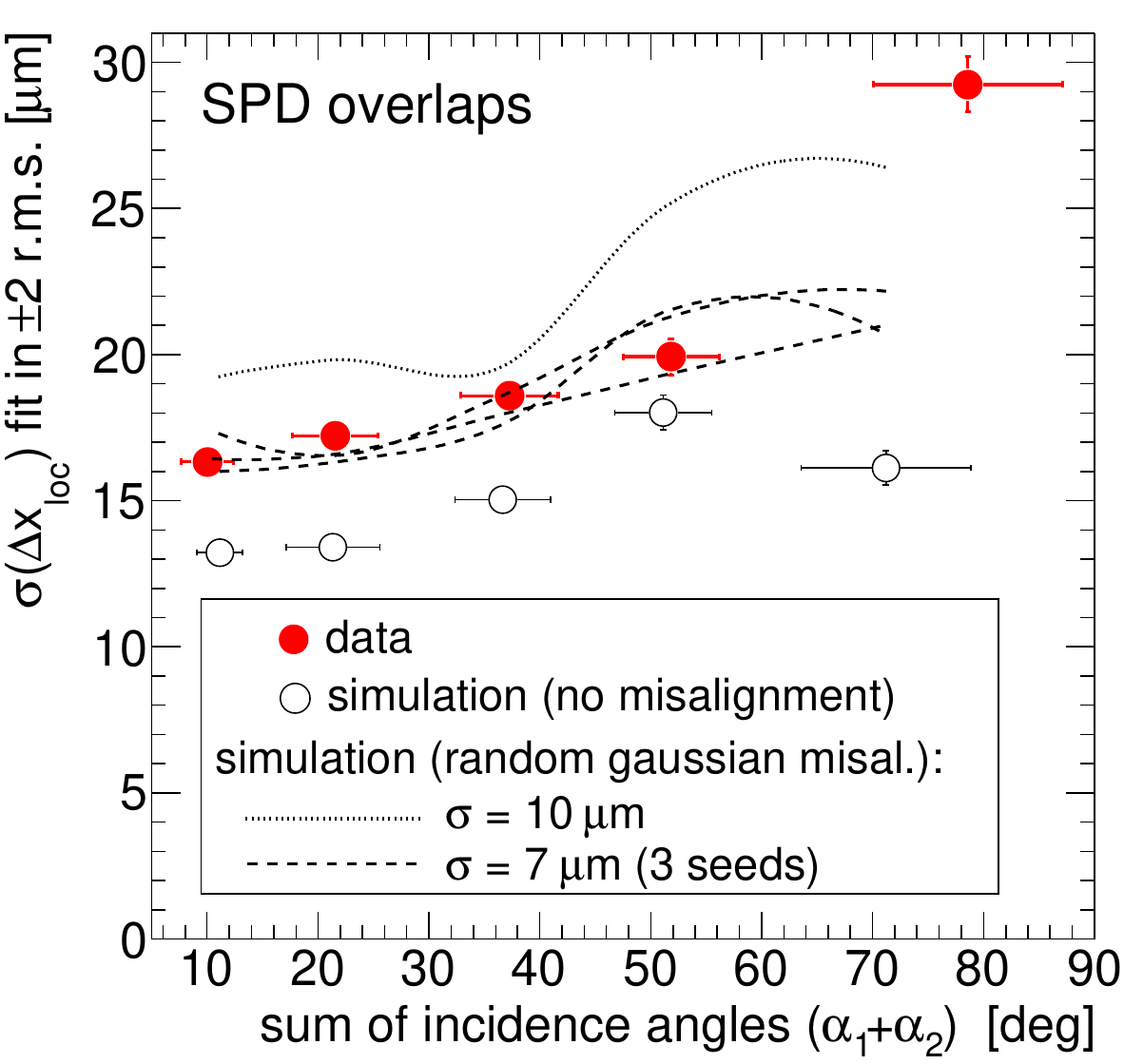}
\caption{(colour online) SPD double points in acceptance overlaps. Left: 
track-to-point $\dxloc$ for ``extra'' points before and after alignment. 
Right: $\sigma$ of the $\dxloc$ distributions as a function of the 
track-to-module incidence angle selection; 2008 cosmic-ray data are 
compared to the simulation with different levels of residual misalignment. 
See text for details.} 
\label{fig:SPDovl}
\end{center}
\end{figure}

We start by showing, in Fig.~\ref{fig:SPDovl} (left), 
the track-to-point distance $\dxloc$ for the SPD ``extra'' points 
in the transverse plane, before and after the Millepede 
alignment. The extra points are not used in the alignment procedure.
The spread of the distribution is $\sigma \approx  18~\mum$, to 
be compared to $\sigma\approx 15~\mum$ from a Monte Carlo simulation with 
ideal geometry. 
An analysis of the $\dxloc$ distance as a function of the $\alpha$ 
incidence angle
 has been performed: 
five windows on the sum
($\alpha_1+\alpha_2$) of the incidence angles on the two 
overlapping modules have been considered. 
These windows define
increasing ranges of incidence angles from $0^\circ$ to $50^\circ$. 
Figure~\ref{fig:SPDovl} (right) shows the spread of the $\dxloc$
distribution for the different incidence angle selections: a clear dependence
of the spread (hence of the spatial resolution) on the incidence angle can be seen. This dependence was already observed in SPD test beam 
measurements~\cite{refSpaResAng1,refSpaResAng2}, which were used to tune the
detector response simulation in the AliRoot software.
In the same figure, Monte Carlo simulation results are reported for comparison:  simulation with ideal geometry (open circles) and with a misaligned geometry obtained using a random gaussian residual
misalignment (dashed lines: misalignments with 
$\sigma=7~\mum$ and three different seeds; dotted line: 
misalignments with $\sigma=10~\mum$).
The 2008 data are well
described by the simulation with a random residual misalignment with 
$\sigma\approx 7~\mum$.
However, this conclusion is based on the assumption that the intrinsic 
resolution is the same in the real detector and in the simulation.
Since the intrinsic resolution can slightly vary depending on the
working conditions of the detector (e.g. the settings used for the bias voltage
and for the threshold), 
the value of $7~\mum$ for the residual misalignment should
be taken only as an indication. Furthermore, this is an equivalent random 
misalignment, while the real misalignments are likely non-gaussian 
and to some extent correlated among different modules.

\begin{figure}[!t]
\begin{center}
\includegraphics[width=0.49\textwidth]{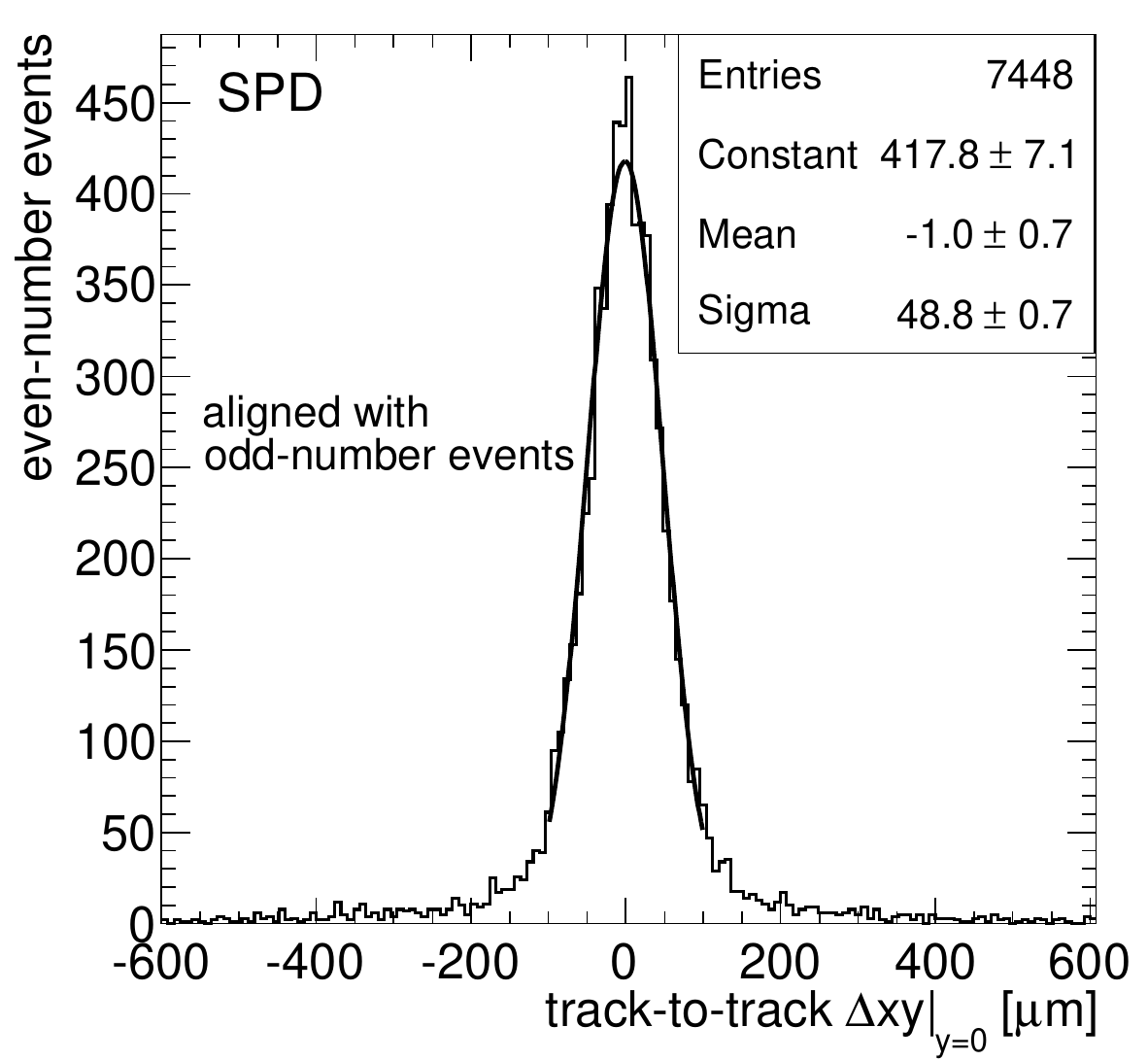}
\includegraphics[width=0.49\textwidth]{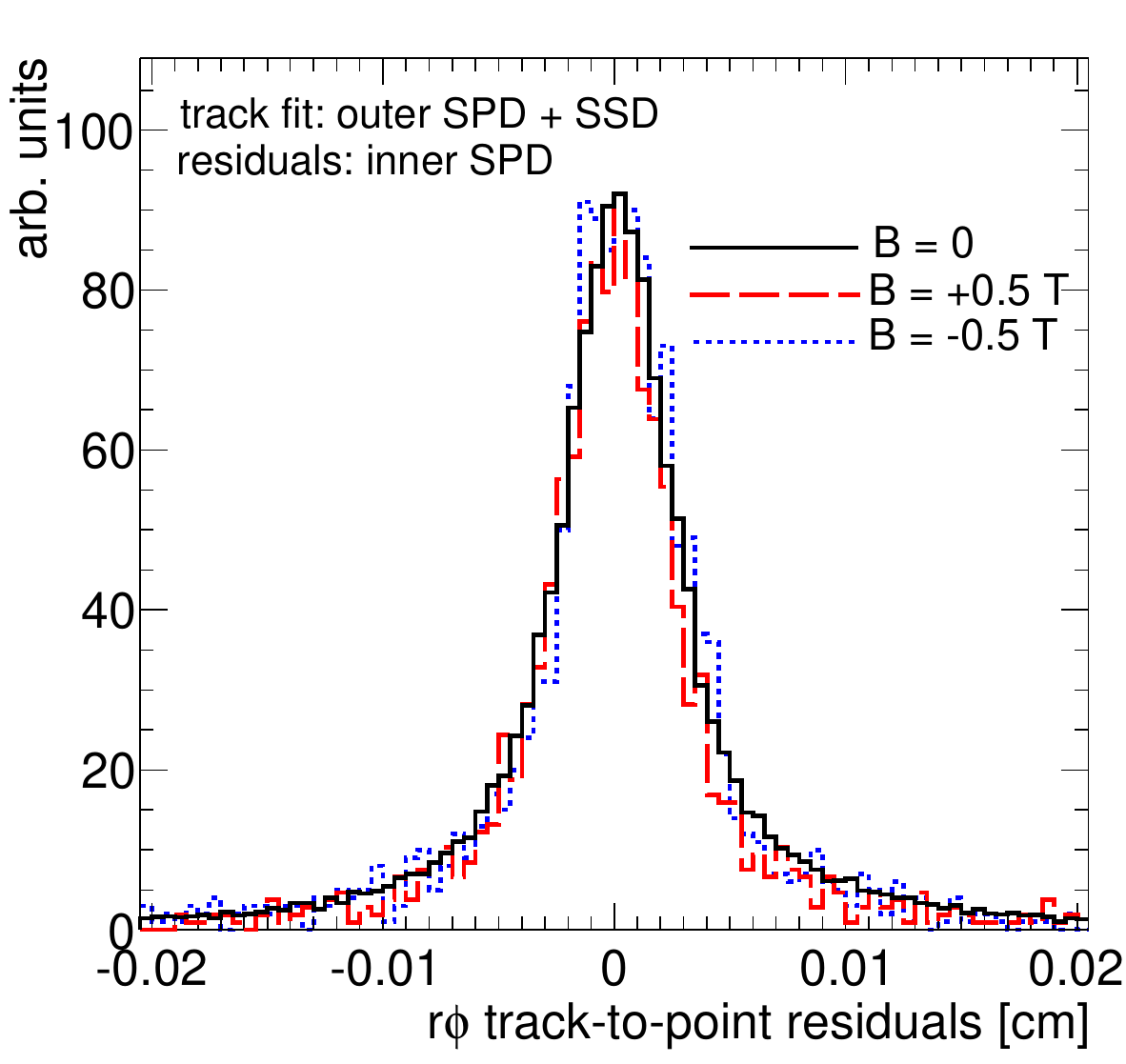}
\caption{(colour online) Alignment stability tests. 
Left: for SPD only, distribution of the $\dxy$ 
distance obtained when aligning with every second track and checking the alignment with the other tracks.
Right: track-to-point residuals in the inner SPD layer (track fit in 
outer SPD layer and in the two SSD layers), for $\rm B=0$ and 
$\rm B=\pm 0.5~T$ (the three histograms 
are normalized to the same integral).} 
\label{fig:SPDstability}
\end{center}
\end{figure}

The robustness of the obtained results was tested by
dividing the data sample in two parts and using every second track
to align the SPD and the others to check the alignment quality.
The corresponding $\dxy$ distribution is presented in the left-hand panel of
Fig.~\ref{fig:SPDstability}:
the distribution is centred at zero and has the same $\sigma\approx50~\mum$
as in the case of aligning with all tracks.

Finally, the data with the 0.5~T magnetic field switched on
(a few thousand events collected
at the end of the 2008 cosmic run) were used to perform dedicated checks to
evaluate a possible effect of the field on the alignment. 
The alignment corrections extracted from data with $\rm B=0$ were applied 
to data with $\rm B=\pm0.5~T$ and 
alignment quality was verified using both the extra points method and 
the track-to-point residuals method.
The distributions of the
track-to-point distance for extra points in SPD acceptance overlaps
for $\rm B=0$, $+0.5$~T and $-0.5$~T data were found to have compatible
widths ($\sigma~[\mum]$: $18.3\pm0.5$, $17.8\pm2.3$, $18.4\pm1.8$, 
respectively)~\cite{noteITSalign}.
Another check was performed using the track-to-point residuals,
calculated by fitting the tracks in the SSD layers and 
the outer SPD layer and evaluating the residuals 
in the inner SPD layer. 
In Fig.~\ref{fig:SPDstability} (right) the comparison between the residuals without magnetic field, with $+0.5$~T, and with $-0.5$~T is shown.
Also in this case, the distributions without field and with the two field 
polarities are compatible.


\subsection{Prospects for inclusion of SDD in the Millepede procedure}

\noindent
The alignment of the SDD detectors
for the $x_{\rm loc}$ coordinate (reconstructed from the drift time)
is complicated by the interplay between the geometrical misalignment 
and the calibration of drift velocity and $t_0$ 
(defined in section~\ref{sec:SDDdescription}).
The $t_0$ parameter accounts for the delays between the time when 
a particle crosses
the detector and the time when the front-end chips receive the trigger 
signal.
Two methods have been developed in order to obtain a first
estimate of the $t_0$ parameter.
The first, and simpler, method consists in extracting the $t_0$
from the minimum measured drift time on a large statistics
of reconstructed SDD points. 
The sharp rising part of the distribution of measured drift times
is fitted with an error function.
The $t_0$ value is then calculated from the fit parameters.
The second method measures the $t_0$ from the 
distributions of residuals along the drift direction ($x_{\rm loc}$) 
between tracks fitted in SPD
and SSD layers and the corresponding points reconstructed in the SDD.
These distributions, in case of miscalibrated $t_0$, show
two opposite-signed peaks corresponding to the two 
separated drift regions of each SDD module, where electrons move in opposite 
directions (see Fig.~\ref{fig:sddstruct}, right).
The $t_0$ can be calculated from the distance of the two peaks and
the drift velocity.
This second procedure has the advantage of requiring smaller statistics,
because it profits from all the reconstructed tracks,
with the drawback of relying on SDD calibration parameters 
(the drift velocity and possibly the correction maps).
Moreover, being based on track reconstruction,
it might be biased by SPD and/or SSD misalignments.


\begin{figure}[!t]
\centering
\resizebox{0.48\textwidth}{!}{%
\includegraphics*[]{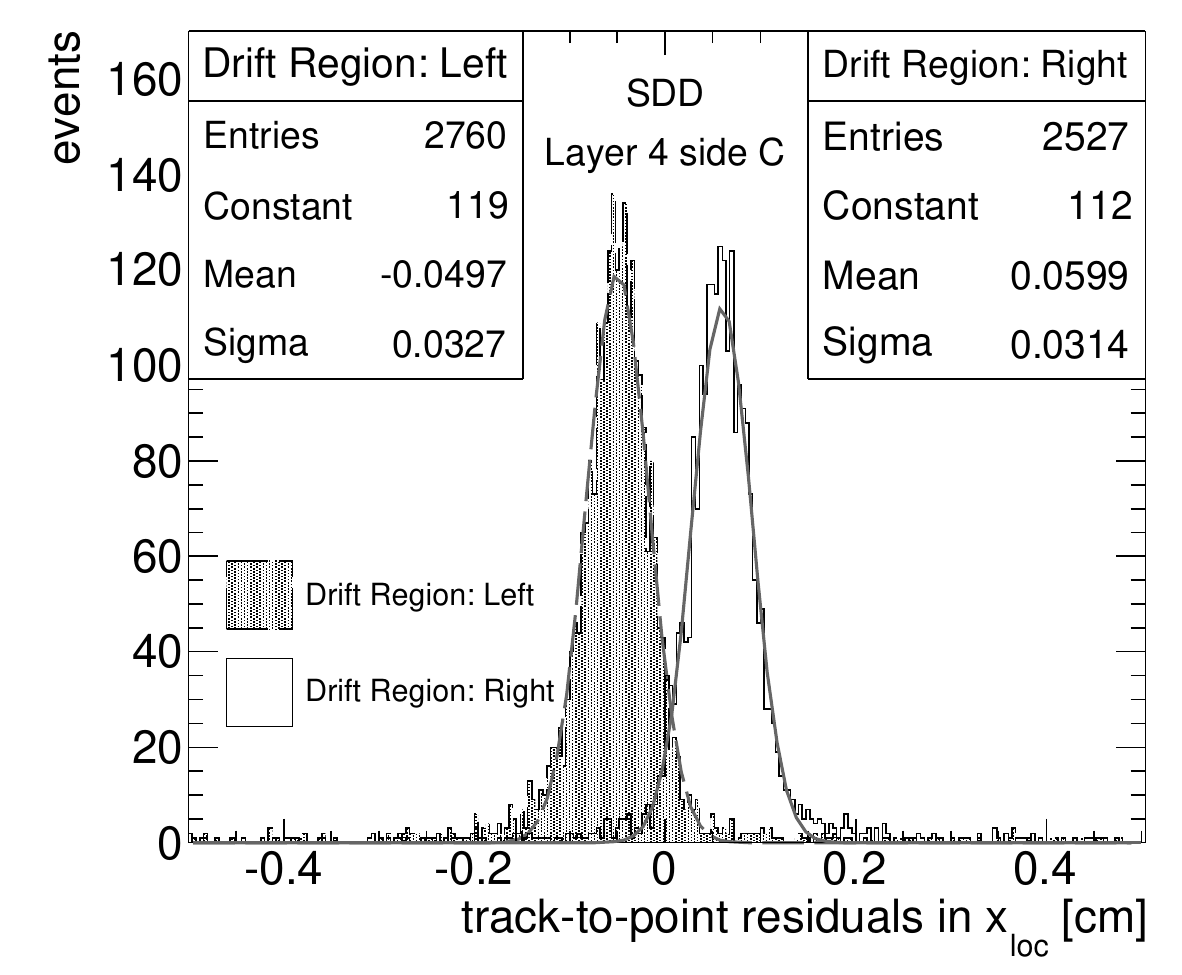}}
\resizebox{0.48\textwidth}{!}{%
\includegraphics*[]{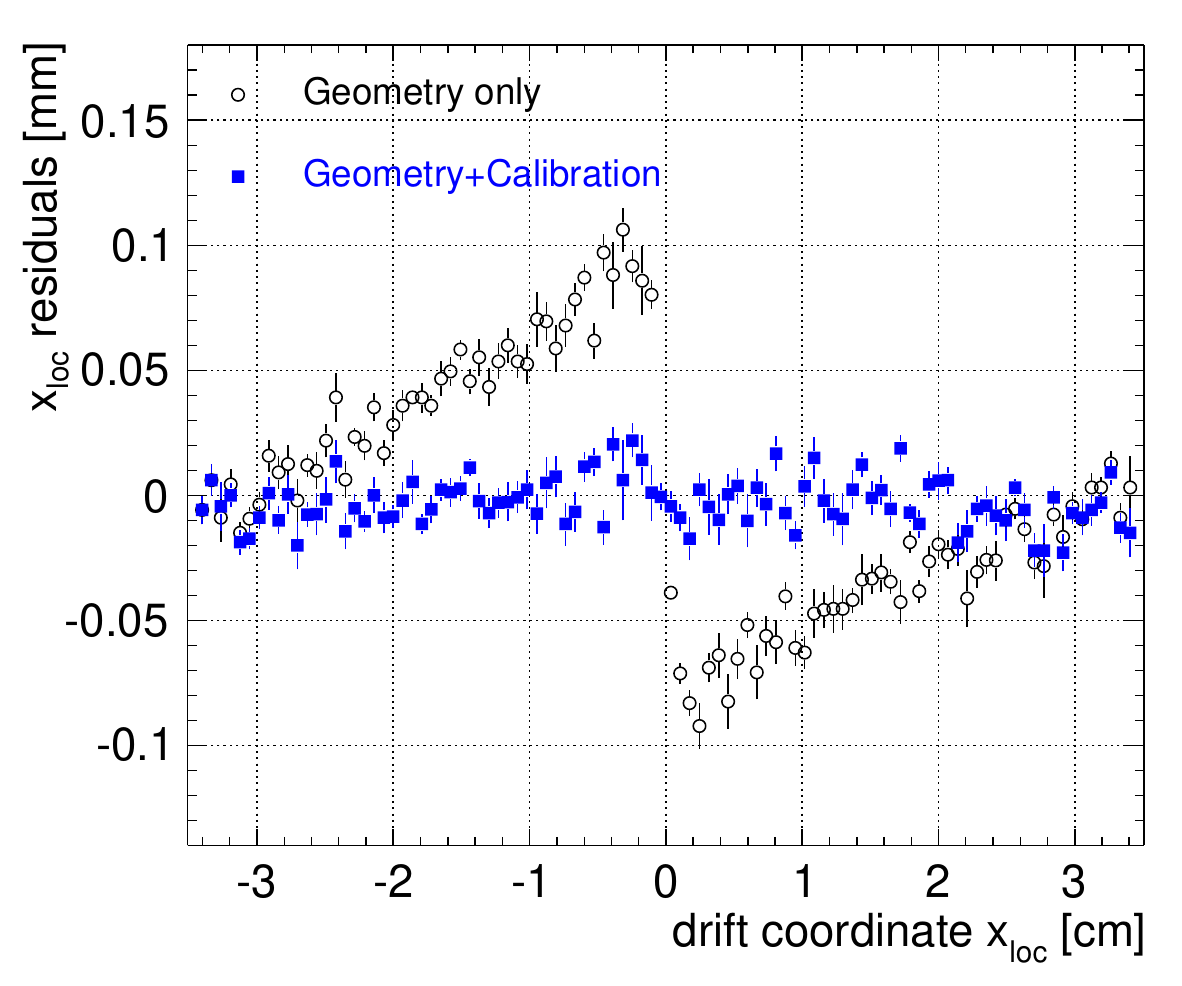}}
\caption{SDD calibration and alignment.
Left: distribution of track-to-point residuals in the two drift regions 
for the SDD modules of layer 4 side C ($z<0$); tracks are fitted using only
their associated points in SPD and SSD; the Millepede alignment 
corrections for SPD and SSD are included, as well as the SSD survey.
Right: residuals along the drift coordinate for one SDD module 
as a function of drift coordinate after Millepede alignment
with only geometrical parameters and with geometrical+calibration
parameters.}
\label{fig:SDDalign}
\end{figure}

Depending on the available statistics, the $t_0$ determination with these two
methods can be done at the level of SDD barrel, SDD ladders or SDD modules.
The $t_0$ parameter needs actually to be calibrated individually
for each of the 260 SDD modules, because of differences in 
the overall length of the 
cables connecting the DAQ cards and the front-end electronics.
In particular, a significant difference is expected
between modules of the A ($z>0$) and C sides ($z<0$), due to the $\approx 6$~m 
difference in the length of the optical fibres connecting the ITS ladders
to the DAQ cards.
With the first 2000 tracks, it is possible to determine the $t_0$
from track-to-point residuals for 4 sub-samples of modules, i.e. separating 
sensors connected to sides A and C of layers 3 and 4.
An example of residual distributions for the left and right drift sides
of the modules of layer 4 side C is shown in 
Fig.~\ref{fig:SDDalign}. The Millepede alignment corrections for SPD
and SSD are applied in this case, and it has been checked that, if they are not applied, the centroid positions in this figure  
are not affected significantly, while the spread of the distributions 
increases, as it could be expected.
A difference of 25~ns between sides A and C of each SDD layer has
been observed, in agreement with the 6~m difference
in fibre lengths (the propagation time of light in optical fibres is
4.89~ns/m).
With larger statistics (35,000 tracks), it is possible to extract 
the $t_0$ for each half-ladder, which requires producing 36(ladders)$\times$2(A/C sides)
pairs of histograms like the ones shown in Fig.~\ref{fig:SDDalign} (left).
Further cable-length differences, which introduce $t_0$ difference, 
exist at the level of individual half ladders and at the level of individual modules. These differences are of the order of 1.5~m (7~ns) and 20~cm ($<1~\rm ns$) respectively, but have not 
been measured in detail yet, because of the limited size and coverage of the cosmic-ray tracks data sample.
It should be noted that given the $\approx6.5~\mum$/ns of drift velocity,
a bias of 1 ns on the $t_0$ can lead to a significant effect on the
reconstructed position along the drift coordinate $x_{\rm loc}$.

After a first calibration with these methods, a refinement 
of the $t_0$ determination is obtained 
by running the Millepede minimization with the $t_0$ as a free global 
parameter for each of the 260 SDD modules.
Similarly, the drift velocity is considered as a free parameter for those SDD modules
(about 35\%) with mal-functioning injectors.
For these modules a single value of drift velocity 
is extracted for the full data sample analyzed, thus neglecting the possible dependence of drift velocity on time due to temperature instabilities.
However, this is not a major concern since the drift velocity was observed to be remarkably stable during the data taking~\cite{sddcomm}.
This allows to assess at the same time geometrical alignment and
calibration parameters of the SDD detectors.
About 500 
tracks are required to align and calibrate a single SDD module. 
An example is shown for a specific SDD module in the
right-hand panel of Fig.~\ref{fig:SDDalign},
where the $x_{\rm loc}$ residuals along the drift direction
are shown as a function of $x_{\rm loc}$.
The result obtained using only the
geometrical rotations and translations as free parameters in the
Millepede minimization is shown by the circle markers.
The clear systematic shift between the two drift regions ($x_{\rm loc}<0$
and $x_{\rm loc}>0$) is 
due to both miscalibrated $t_0$ and biased drift velocity
(this is a module with non-working injectors).
These systematic effects are no longer present when 
also the calibration parameters are
fitted by Millepede (square markers). 
It should be pointed out that the width of the SDD residual distributions
shown in Fig.~\ref{fig:SDDalign}
does not correspond to the expected resolution on SDD points along drift
coordinate because of jitter between the time when the muon crosses
the detectors and the SPD FastOR trigger, which has an
integration time of 100~ns.
For the about 100 SDD modules with highest occupancy, the statistics collected in the 2008 cosmic run allowed to check the reliability of the calibration parameters ($t_0$ and drift velocity) extracted with Millepede by comparing the values obtained from independent analyses of two sub-samples of tracks.
From this study, a precision of $0.025~\mum$/ns for the drift velocity 
and 10~ns for $t_0$ was estimated. It should be noted that 
these precisions are limited by the available statistics as well as by the 
trigger jitter effect mentioned above.


\section{SPD alignment with an iterative local method}
\label{sec:iterative}

\noindent
We developed an alignment method
that performs a (local) minimization for each single module 
and accounts for correlations between modules by iterating the procedure until convergence is reached.
A similar approach is considered by both the CMS and ATLAS 
experiments~\cite{refIterAtlas,refIterCMS,refIterCMS_Kari}.
The main difference between this method and the Millepede algorithm is that only
in the latter the 
correlations between the alignment parameters of all modules are 
explicitly taken into account.
Conversely, the local module-by-module algorithm assumes
that the misalignments of the modules crossed by a given track are uncorrelated
and
performs the minimization of the residuals independently for each module. 
The comparison of the alignment parameters
from this method and from Millepede
would provide
a further validation of the results achieved with the Millepede.

In the local method we minimize, module-by-module, 
the following local $\chi^2$ function of the
alignment parameters of a single module:
\begin{eqnarray}
\label{eq:iterative}
\chi^2_{\rm local}({\vec a}_{\rm tra},{\bf A}_{\rm rot}) & = &
\sum_{\rm tracks}{\vec{\delta}_{t,p}^{\rm\,T} \, {\bf V}^{-1}_{t,p} \, \vec{\delta}_{t,p}}\nonumber \\ 
&=&\sum_{\rm tracks}\left({\vec r}_{t}-{\bf A}_{\rm rot}{\vec r}_{p}-{\vec a}_{\rm tra}\right)^{\rm T}\left({\bf V}_{t}+{\bf V}_{p}\right)^{-1}\left({\vec r}_{t}-{\bf A}_{\rm rot}{\vec r}_{p}-{\vec a}_{\rm tra}\right)\,.
\end{eqnarray}
Here, the sum runs over the tracks passing through the module, 
${\vec r}_{p}$ is the position of the 
measured point on the module while ${\vec r}_{t}$ is the crossing point 
on the module plane
of the track $t$ fitted with all points but ${\vec r}_{p}$.
${\bf V}_{t}$ and ${\bf V}_{p}$ are the covariance matrices of the crossing point and of the measured point, respectively. 
The six alignment parameters enter this formula in the vector ${\vec a}_{\rm tra}$,
the alignment correction for the position of the centre of the module,
and in the rotation matrix ${\bf A}_{\rm rot}$, the alignment 
correction for the orientation of the plane of the module. 
The alignment correction is supposed to be small so that the rotation matrix can be approximated
as the unity matrix plus a matrix linear in the angles. In this way, 
the $\chi_{\rm local}^2$ is a quadratic expression
 of the alignment parameters and the minimization can be performed by simple inversion.
The $\chi_{\rm local}^2$ function in Eq.~(\ref{eq:iterative}) can be written in the same way also for a 
set of modules considered as a rigid block.
The track parameters are not affected by the misalignment of the module under
study, because the track point on this module is not used in the fit, while
the positions of the crossing points are affected, 
because the tracks are propagated to the
plane of the module defined in the ideal geometry. 
This is taken into account by adding 
a large error along the track direction to the covariance matrix of the 
crossing point.

Given that this is a local method, it is expected to work best if 
two conditions are fulfilled: the correlation between the misalignments of different modules is small and
the tracks used to align a given module cross several 
other modules. In order to limit the bias that can be 
introduced by modules with low statistics, for which the second condition 
is normally not met, we align
the modules following a sequence of decreasing number of points.
To reduce the residual correlation between the alignment parameters 
obtained for the different modules, we iterate the procedure
until the parameters converge. Simulation studies with misalignments
of the order of $100~\mum$ have shown that the
 convergence is reached after about 10 iterations.

\begin{figure}[!t]
  \begin{center}
    \includegraphics[width=.49\textwidth]{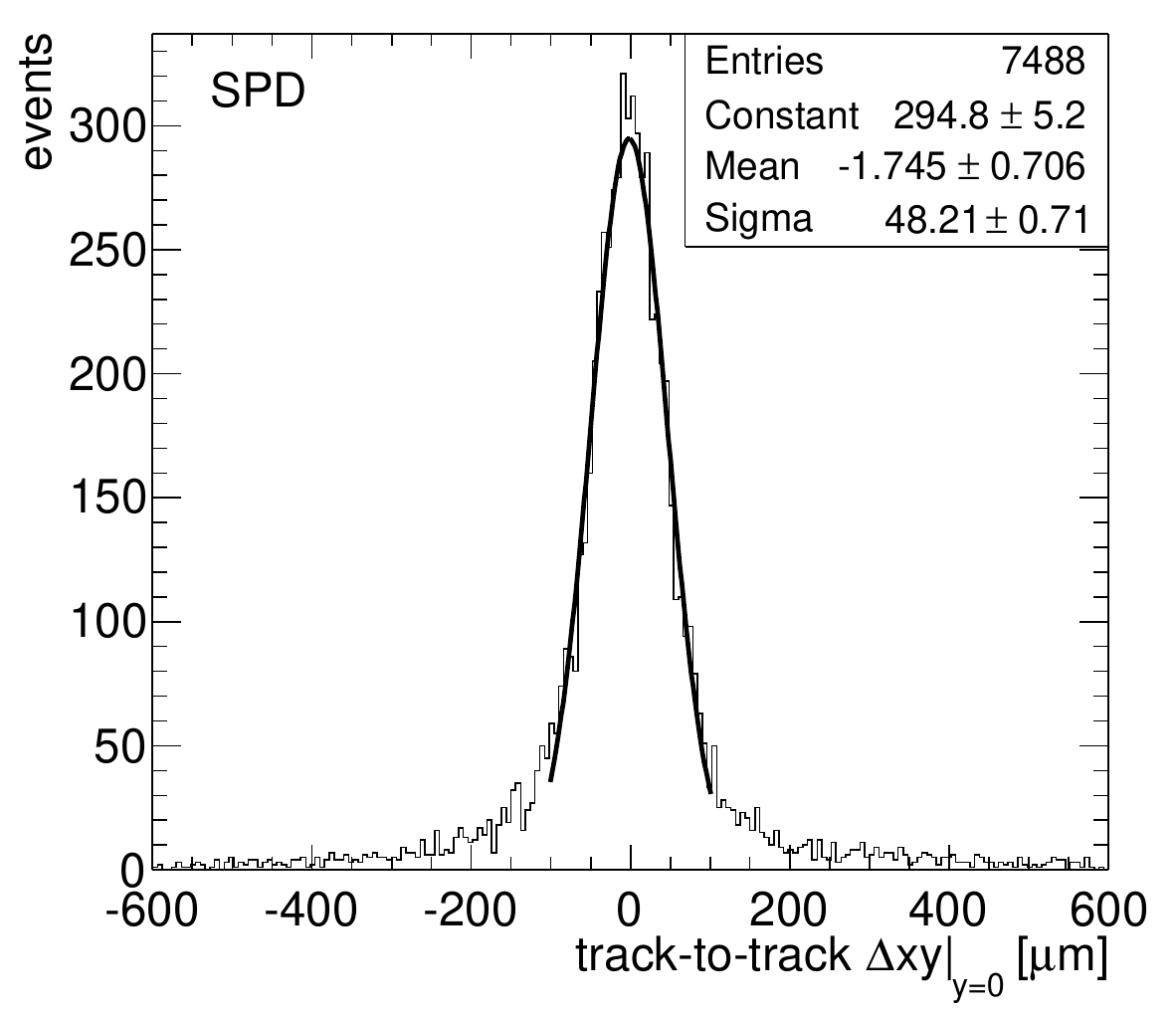}
    \includegraphics[width=.49\textwidth]{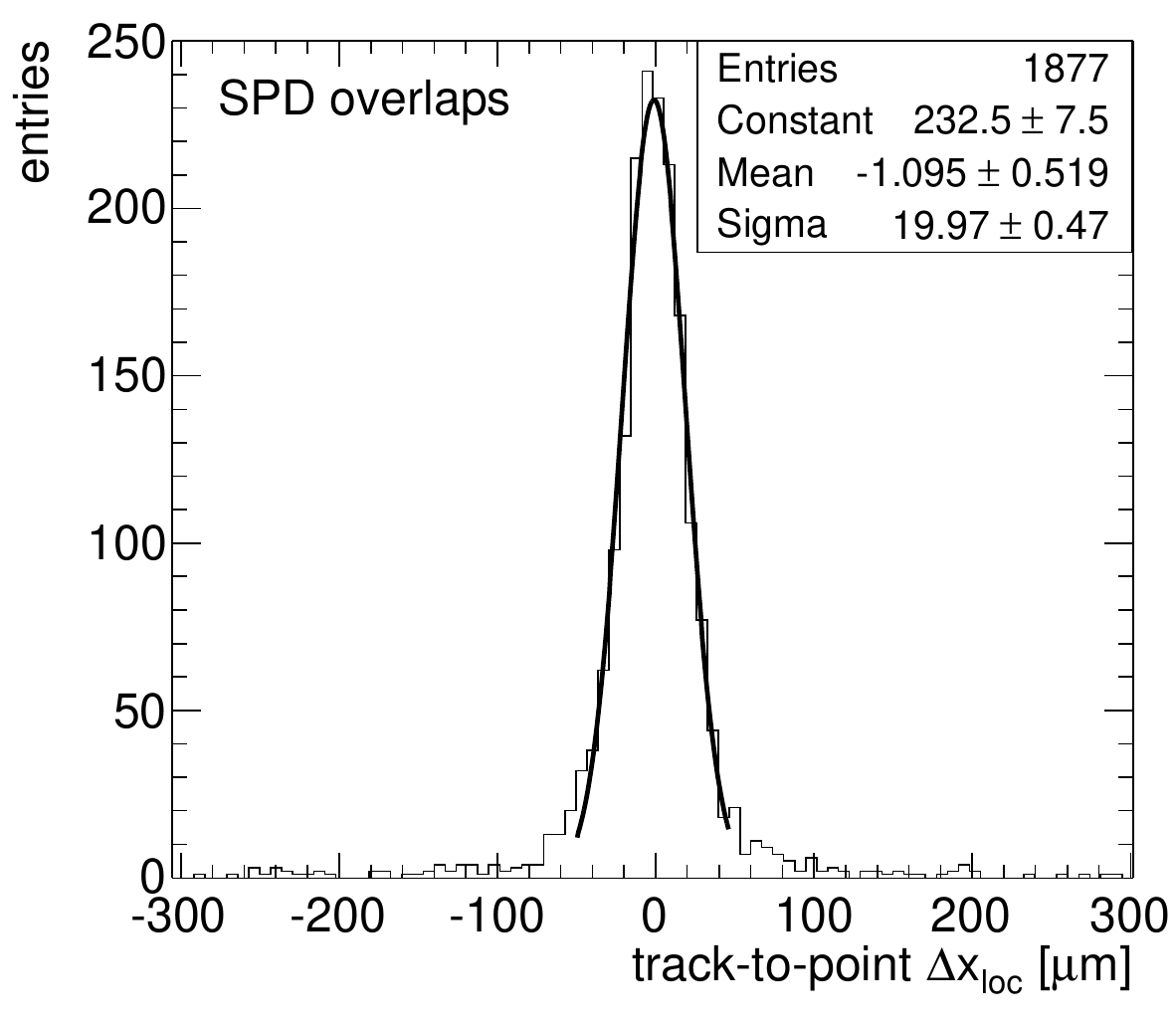}
    \caption{SPD alignment quality results for the iterative local method.
     Left: track-to-track $\dxy$ distribution defined 
    in section~\ref{sec:milleresults}, using only SPD points.
       Right: track-to-point $\dxloc$ 
        distribution for extra points in acceptance overlaps.}
      \label{fig:IterDXYatY0}
  \end{center}
\end{figure}

\begin{figure}[!t]
  \begin{center}
    \includegraphics[width=.98\textwidth]{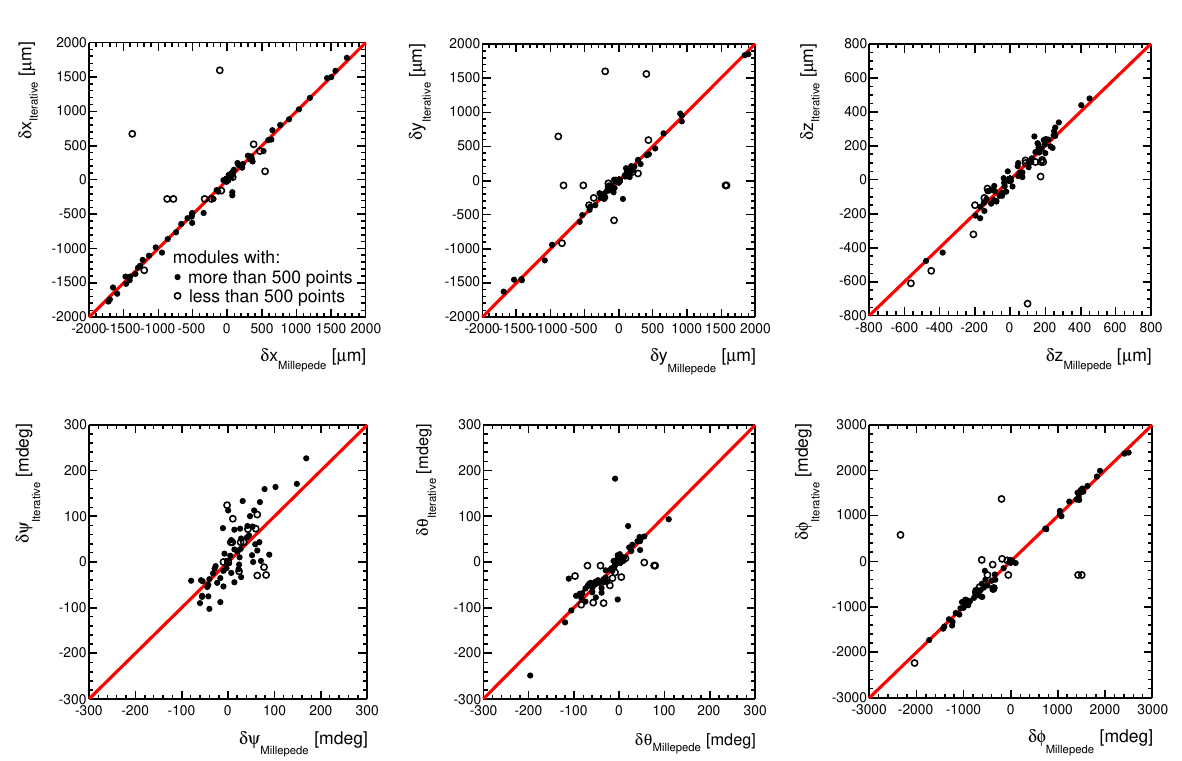}
    \caption{Correlation between the alignment parameters obtained from Millepede
    (horizontal axis) and the iterative method (vertical axis), 
    for the inner SPD layer modules. Modules with more (less) than 500 points 
    are represented by the closed (open) markers.}
    \label{fig:CompareScattSPD1}
  \end{center}
\end{figure}

For the ITS alignment using the 2008 cosmic-ray 
data, we aligned only the SPD modules
using this method. Like for Millepede, we adopted a hierarchical
 approach. Given the excellent precision of the SSD survey measurements,
we used these two layers as a reference. We aligned as a first step
the whole SPD barrel with respect to the SSD, 
then the two half-barrels with respect to
the SSD, then the SPD sectors with respect to the SSD. In the last step, we used SPD and SSD points to fit the tracks and we aligned the individual 
sensor modules of the SPD.
Figure~\ref{fig:IterDXYatY0} shows the top--bottom track-to-track $\dxy$ 
distribution obtained using only the SPD points (left-hand panel) and the 
track-to-point $\dxloc$ for the double points in acceptance overlaps
(right-hand panel), after alignment. Both distributions are compatible (mean and
sigma from a gaussian fit) with the corresponding distributions after
Millepede alignment.
This is an important independent verification of the Millepede results.
Since the two methods are in many aspects independent, comparing the
two sets of alignment parameters could provide a check for 
the presence of possible systematic trends. 
Figure~\ref{fig:CompareScattSPD1} shows the correlation of the inner SPD 
layer parameter values
obtained
 with the iterative method and those obtained with Millepede. 
A correction
was applied to account for a possible global roto-translation of the whole ITS, 
which does not affect the quality of the alignment and can be different for the two methods. 
The closed (open) markers represent the modules with more (less) than 500
track points.
Most of the modules are clustered along the diagonal lines where 
the parameters from the two methods 
are exactly the same. There are some outlier modules, that are
far from the ideal result. However, these outliers mostly correspond
to modules with low statistics (open markers) at the sides of the SPD barrel.


\section{Conclusions}
\label{sec:conclusions}

\noindent
The results on the first alignment of the ALICE Inner Tracking System with 
cosmic-ray tracks, collected in 2008 in the absence of magnetic field, have been 
presented. 

The initial step of the alignment procedure consisted of 
the validation of the survey measurements for the Silicon Strip Detector (SSD).
The three methods applied for this purpose indicate that the residual
misalignment spread for modules on ladders is less than $5~\mum$, i.e. negligible 
with respect to the intrinsic resolution of this detector in the most precise
direction, while the
residual misalignment spread for the ladders with respect to the support cones 
amounts to about $15~\mum$.

The procedure continues with track-based software alignment
performing residuals minimization. We presented the results obtained 
with a sample of about $10^5$ cosmic-ray tracks, reconstructed in events
selected by the FastOR trigger of the Silicon Pixel Detector (SPD). 
We mainly use the Millepede algorithm, which minimizes a global 
$\chi^2$ of residuals for all alignable volumes and a large set of tracks.

We start from the SPD, 
which is aligned in a hierarchical 
approach, from the largest mechanical structures (10 support sectors) to 
the 240 single sensor modules. About 90\% of the latter were active during
the 2008 cosmic run, and about 85\% had enough space points 
($>50$) to perform alignment.
Then, we align the SPD barrel with respect to the 
SSD barrel. The SSD coverage provided by the cosmic-ray tracks is insufficient 
to align the SSD at the level of ladders, especially for the ladders
close to the horizontal plane $y=0$. 
Therefore, for the time being 
we only align the SSD at the level of large sets of ladders. 

The two 
intermediate ITS layers, the Silicon Drift Detectors (SDD), represent
 a special case, 
 because the reconstruction of one of the two local coordinates 
requires dedicated calibration procedures (drift velocity and drift time zero
extraction), which are closely related to the alignment. Indeed, 
one of the approaches that we are developing for the time zero calibration
is based on the analysis of track residuals in a standalone procedure, initially, 
and then directly within the Millepede algorithm. Once these procedures
are stable and robust, the SDD will be included in the standard 
alignment chain.
For all six layers, the completion of the alignment for all modules will 
require tracks from proton--proton collisions;
 a few $10^6$ events (collected in a few days)
should allow us to reach a uniform alignment level, close
to the target, over the entire detector.

We use mainly two observables to assess the quality 
of the obtained alignment: the matching of the two half-tracks produced 
by a cosmic-ray particle in the upper and lower halves of the ITS barrel,
and the residuals between double points produced in the geometrical overlaps
between adjacent modules. For the SPD, both observables indicate 
an effective space point resolution of about $14~\mum$ in the most precise
direction, only 25\% worse than the resolution of 
about $11~\mum$ extracted from the Monte Carlo
simulation without misalignments. 
In addition, the measured incidence angle
dependence of the spread of the double points residuals is well 
reproduced by Monte Carlo simulations that include random residual misalignments
with a gaussian sigma of about 7~$\mum$. Further confidence on the 
robustness of the results is provided, to some extent, 
by the cross-checks we performed
using a small data set with magnetic field switched 
on and, mainly, by the comparison of
the Millepede results to those from a second, independent, alignment method.
This second method, which iteratively minimizes a set of local
module-by-module $\chi^2$ functions, yields, compared to Millepede, 
a similar alignment quality and a 
quite compatible set of alignment corrections.

Using the present data set with magnetic field off, since the track momenta
are not known, the multiple scattering
effect, which is certainly not negligible, cannot be disentangled from 
the residual misalignment effect. Therefore, a more conclusive statement
on the SPD residual misalignment 
will be possible only after the analysis of cosmic-ray data collected 
with magnetic field switched on. 
The same applies for combined tracking with SPD, SDD and SSD points: 
in this case, the
momentum-differential analysis of the transverse distance between the two
half-tracks (upper and lower half-barrels) will allow us to measure the 
track transverse impact parameter resolution, which is a key
performance figure in view of 
the ALICE heavy flavour physics program.



\section*{Acknowledgements}

The ALICE collaboration would like to thank all its engineers and technicians for their invaluable contributions to the construction of the experiment,
and, in particular, of the Inner Tracking System.

The ALICE collaboration acknowledges the following funding agencies for their support in building and
running the ALICE detector:
\begin{itemize}
\item{}
Calouste Gulbenkian Foundation from Lisbon and Swiss Fonds Kidagan, Armenia;
\item{}
Conselho Nacional de Desenvolvimento Cient'fico e Tecnol—gico (CNPq), Financiadora de Estudos e Projeto (FINEP),
Funda\c{c}\~{a}o de Amparo \`{a} Pesquisa do Estado de S\~{a}o Paulo (FAPESP);
\item{}
National Natural Science Foundation of China (NSFC), the Chinese Ministry of Education (CMOE)
and the Ministry of Science and Technology of China (MSTC);
\item{}
Ministry of Education and Youth of the Czech Rebublic;
\item{}
Danish National Science Research Council and the Carlsberg Foundation;
\item{}
The European Research Council under the European Community's Seventh Framework Programme;
\item{}
Helsinki Institute of Physics and the Academy of Finland;
\item{}
French CNRS-IN2P3, the `Region Pays de Loire', `Region Alsace', `Region Auvergne' and CEA, France;
\item{}
German BMBF and the Helmholtz Association;
\item{}
Hungarian OTKA and National Office for Research and Technology (NKTH);
\item{}
Department of Atomic Energy and Department of Science and Technology of the Government of India;
\item{}
Istituto Nazionale di Fisica Nucleare (INFN) of Italy;
\item{}
MEXT Grant-in-Aid for Specially Promoted Research, Ja\-pan;
\item{}
Joint Institute for Nuclear Research, Dubna;
\item{}
Korea Foundation for International Cooperation of Science and Technology (KICOS);
\item{}
CONACYT, DGAPA, M\'{e}xico, ALFA-EC and the HELEN Program (High-Energy physics Latin-American--European Network);
\item{}
Stichting voor Fundamenteel Onderzoek der Materie (FOM) and the Nederlandse Organistie voor Wetenschappelijk Onderzoek (NWO), Netherlands;
\item{}
Research Council of Norway (NFR);
\item{}
Polish Ministry of Science and Higher Education;
\item{}
National Authority for Scientific Research - NASR (Autontatea Nationala pentru Cercetare Stiintifica - ANCS);
\item{}
Federal Agency of Science of the Ministry of Education and Science of Russian Federation, International Science and
Technology Center, Russian Federal Agency of Atomic Energy, Russian Federal Agency for Science and Innovations and CERN-INTAS;
\item{}
Ministry of Education of Slovakia;
\item{}
CIEMAT, EELA, Ministerio de Educaci\'{o}n y Ciencia of Spain, Xunta de Galicia (Conseller\'{\i}a de Educaci\'{o}n),
CEA\-DEN, Cubaenerg\'{\i}a, Cuba, and IAEA (International Atomic Energy Agency);
\item{}
Swedish Reseach Council (VR) and Knut $\&$ Alice Wallenberg Foundation (KAW);
\item{}
Ukraine Ministry of Education and Science;
\item{}
United Kingdom Science and Technology Facilities Council (STFC);
\item{}
The United States Department of Energy, the United States National
Science Foundation, the State of Texas, and the State of Ohio.
\end{itemize}


\end{document}